\documentclass[%
 aip,
 amsmath,amssymb,
 reprint,%
]{revtex4-2}

\usepackage{graphicx}
\usepackage{dcolumn}
\usepackage{bm}

\usepackage[utf8]{inputenc}
\usepackage[T1]{fontenc}
\usepackage{mathptmx}
\usepackage{etoolbox}
\usepackage{color}
\usepackage{xcolor}
\usepackage{bigints}

\newcommand{\gpt}{\dot{\gamma}}
\newcommand{\p}{{\prime}}
\newcommand{\pp}{{\prime \prime}}
\usepackage{comment} 
\usepackage{hyperref}

\makeatletter
\def\@email#1#2{%
 \endgroup
 \patchcmd{\titleblock@produce}
  {\frontmatter@RRAPformat}
  {\frontmatter@RRAPformat{\produce@RRAP{*#1\href{mailto:#2}{#2}}}\frontmatter@RRAPformat}
  {}{}
}%
\makeatother
\begin{document}

\preprint{AIP/123-QED}

\title[]{Aging dynamics and generalized superposition principles in carbon~black-carboxymethylcellulose hydrogels}
\author{Gauthier Legrand}
\affiliation{ENSL, CNRS, Laboratoire de physique, F-69342 Lyon, France}
\author{Guilhem P. Baeza}%
\affiliation{Universit\'e Claude Bernard Lyon 1, INSA Lyon, Universit\'e Jean Monnet, CNRS, UMR 5223, Ing\'enierie des Mat\'eriaux Polymères, F-42023 Saint-Etienne Cedex, France}
\author{Jarno L. Waeterloos}
\affiliation{Department of Chemical Engineering, KU Leuven, Celestijnenlaan 200J, Box 2424, 3001 Leuven, Belgium}
\author{Christian Clasen}
\affiliation{Department of Chemical Engineering, KU Leuven, Celestijnenlaan 200J, Box 2424, 3001 Leuven, Belgium}
\author{S\'ebastien Manneville}
\affiliation{ENSL, CNRS, Laboratoire de physique, F-69342 Lyon, France}
\affiliation{Institut Universitaire de France (IUF)}
\author{Thibaut Divoux}
\affiliation{ENSL, CNRS, Laboratoire de physique, F-69342 Lyon, France}

\date{\today}

\begin{abstract}
Aqueous dispersions of hydrophobic carbon black (CB) particles can be stabilized by carboxymethylcellulose (CMC), forming CB-CMC hydrogels whose microstructure is controlled by the relative CB-to-polymer content. At low CMC content, CB particles form a percolated, electrically conductive network, whereas at higher polymer content, CB particles act as isolated physical crosslinkers within a an electrically insulating polymer matrix [see Legrand \textit{et al.}, Macromolecules 56, 2298-2308 (2023)]. 
Combining rheology and electrical impedance spectroscopy, we show that these two regimes exhibit markedly different aging dynamics. Conductive hydrogels exhibit logarithmic elastic recovery associated with a percolated CB network, whereas insulating hydrogels exhibit a combination of logarithmic and power-law stiffening, accompanied by dielectric polarization associated with confined ionic motion. The insulating hydrogels further exhibit a time-aging superposition principle, as their viscoelastic spectra measured in a time-resolved fashion can collapse onto a single master curve. 
Under steady shear, the viscoelastic response measured perpendicular to the flow also obeys a time--shear-rate superposition principle. Remarkably, the scaling between the shift factors under shear is reminiscent of critical-gel behavior, unlike that observed during aging at rest.
Finally, time-resolved rheo-USAXS measurements performed during shear-rate cycles reveal structural rearrangements of CB particles and clusters that correlate with rheological hysteresis. These results provide a microstructural basis for understanding the nonlinear rheology of insulating CB-CMC hydrogels and establish generalized superposition principles as a unifying framework for their aging and flow behavior.
\end{abstract}

\maketitle

\section{\label{sec:intro} Introduction}
Physical aging is the hallmark of numerous amorphous soft solids, including colloidal gels, polymer networks, and dense suspensions.\cite{Fielding:2000,Bouzid:2017,Bonn:2017,Joshi:2018} Following a mechanical perturbation, these systems progressively evolve toward increasingly constrained microstructures, resulting in a slow increase in the elastic properties and continuous shift of their viscoelastic spectrum. Such dynamics are commonly reflected in logarithmic or power-law growth of the elastic modulus,\cite{TeNijenhuis:1975,Derec:2003,Manley:2005,Coussot:2006,Guo:2011,Aime:2018,Doria:2023,Bauland:2024} as well as the existence of time-aging (or time-connectivity) superposition principles, whereby viscoelastic spectra measured at different waiting times can be collapsed onto a single master curve through appropriate rescaling.\cite{Ramos:2001,Shaukat:2010,Shahin:2012,Jiang:2018,Keshavarz:2021,MorletDecarnin:2023,MorletDecarnin:2026} These observations have proved instrumental in characterizing the evolution of soft materials at rest and in providing a compact scaling description of their evolving mechanical response. 

Related scaling concepts have also been reported under flow using orthogonal superposition rheometry,\cite{Vermant:1997} where viscoelastic spectra measured under steady shear, and perpendicularly to the flow direction, can be collapsed onto master curves through a time--shear-rate superposition.\cite{Sung:2018,Lee:2021} Whether such superposition principles observed at rest and under flow are manifestations of a common underlying structural evolution, however, remains largely unexplored.

In this context, aqueous dispersions of carbon black (CB) particles stabilized by carboxymethylcellulose (CMC) constitute a versatile model system in which the nature of the aging can be tuned simply by varying the composition. Depending on the relative CB-to-CMC content, these hydrogels exhibit two distinct microstructural regimes.\cite{Legrand:2023} At low polymer concentration, CB particles form a percolated network stabilized by polymer adsorption, resulting in electrically conductive gels (conductivity of about\cite{Legrand:2023} $3~\rm mS.cm^{-1}$) whose properties are governed by the connectivity of the CB network. At higher polymer content, CB particles instead act as physical crosslinkers within a continuous polymer matrix, yielding electrically insulating gels whose mechanical response is governed by a polymer network with transient CB-mediated junctions. These two regimes are separated by a well-defined depercolation transition of the CB network, which occurs at a critical CMC-to-CB mass ratio, written $r_c$, and marks the loss of long-range electrical connectivity.\cite{Legrand:2023}
These two types of colloid-polymer gels provide an ideal model system for comparing how distinct microstructures age and recover following mechanical rejuvenation.

In the first part of this work, we use rheology and electrical impedance spectroscopy to compare the aging dynamics of conductive and insulating CB-CMC hydrogels. While conductive gels display logarithmic rheological recovery dynamics, as classically reported in soft glassy materials,\cite{Joshi:2018} insulating gels exhibit a two-step recovery with an initial logarithmic growth at short aging times followed by a power-law increase at longer times, suggesting that distinct microscopic processes successively dominate the aging dynamics. 
Electrical impedance spectroscopy further reveals fundamentally different electrical relaxation processes associated with the two microstructures. In conductive gels, the electrical response associated with aging is dominated by charge transport through the percolated CB network. In contrast, insulating CB-CMC hydrogels exhibit dielectric relaxation arising from the accumulation of mobile ions at interfaces within the polymer-rich matrix (Maxwell-Wagner-Sillars-type polarization), reflecting the absence of a continuous electronically conducting pathway.
Because of their comparatively slow recovery dynamics, insulating CB-CMC hydrogels provide an ideal model system for investigating the time-resolved evolution of the linear viscoelastic spectrum during aging.

Building on these observations, the remainder of this work focuses on insulating CB-CMC hydrogels to investigate whether the structural evolution observed during aging at rest can be quantitatively linked to the material response under shear. Using time-resolved mechanical spectroscopy, we first show that the viscoelastic spectra of insulating CB-CMC hydrogels obey a time-aging superposition principle and are well described by a Fractional Kelvin-Voigt model, extending the time-concentration superposition previously reported for these systems at rest.\cite{Legrand:2023}
We then investigate whether an analogous scaling framework persists under steady shear using orthogonal superposition rheometry, which probes the linear viscoelastic response perpendicular to the flow.\cite{Vermant:1997} 
Remarkably, the spectra measured under shear also collapse onto a single master curve after appropriate rescaling, revealing a generalized time-shear-rate superposition up to a critical shear rate beyond which the elastic contribution vanishes.
Finally, we investigate the nonlinear rheology of insulating CB-CMC hydrogels during decreasing and increasing shear-rate sweeps.
By combining rheology with time-resolved ultra-small-angle X-ray scattering (TRUSAXS), we identify shear-induced rearrangements of CB clusters acting as transient crosslinks within the polymer network. These measurements provide a microscopic picture of the structural evolution underlying both the elastic response revealed by orthogonal superposition rheometry and the rheological hysteresis observed under continuous shear. 

The article is organized as follows. Section~\ref{sec:materials} describes the sample preparation and the techniques employed to study CB-CMC hydrogels. In section~\ref{sec:aging}, we compare the mechanical and electrical aging of conductive and insulating CB-CMC hydrogels at rest using rheometry and electrical impedance spectroscopy. We then focus on insulating hydrogels to establish a time-aging superposition principle, while in Section~\ref{sec:flow}, we further identify a time--shear-rate superposition principle from linear viscoelastic measurements performed by means of orthogonal superposition under steady shear. Finally, rheo-TRUSAXS measurements performed during shear-rate cycles provide direct structural insight into the evolution of the CB clusters acting as transient crosslinks within the polymer matrix. These measurements establish a microscopic connection between the structural rearrangements occurring under flow, the emergence of elasticity captured by orthogonal superposition rheometry, and the rheological hysteresis observed during shear-rate cycles. Section~\ref{sec:discussion} offers a compact discussion and concluding remarks.

\section{\label{sec:materials} Materials and methods}

\subsection{Sample preparation}

CB–CMC hydrogels were prepared starting from aqueous solutions of sodium carboxymethylcellulose (NaCMC, Sigma-Aldrich, ref. 419303). The polymer has a nominal molecular weight $M_w=250~\rm kg.mol^{-1}$ and a degree of substitution $DS=0.9$; independent characterization yielded $M_w=213~\rm kg.mol^{-1}$ and $DS=0.88$ (see Ref.~\onlinecite{Legrand:2023}). NaCMC was dissolved in deionized water to prepare stock solutions with polymer concentrations up to 5 wt.\%. Dissolution was achieved by gentle mixing on a tube roller at room temperature for 48~h, until complete homogenization.

Carbon black (CB) particles (VXC72R, Cabot Corp.) were subsequently incorporated into the NaCMC solutions. These particles are hydrophobic colloids composed of quasi-spherical primary units with a typical diameter of $20~\rm nm$, irreversibly aggregated into branched, fractal particles of characteristic size of about $500~\rm nm$.\cite{Richards:2017,Legrand:2026} To ensure proper dispersion, CB-CMC mixtures were subjected to two ultrasonic bath treatments of 90~min each, separated by 24~h of magnetic stirring. Following sonication, samples were left to rest for an additional 24~h prior to experiments.

Throughout this work, samples are identified by the CMC concentration $c_{\rm CMC}$, expressed in wt.\% relative to the CB-free stock solution, and by the CB mass fraction $x_{\rm CB}$ in the final dispersion, following the notation introduced in Ref.~\onlinecite{Legrand:2023}.

\subsection{Electrical Impedance Spectroscopy}

Electrical impedance spectroscopy was performed in AC mode on a sample at rest loaded in a cylindrical cell made of Teflon (thickness of about $1$ mm and surface of about 0.5~cm$^2$). The inner sides were made of metal and acted as electrodes that were connected to an impedance analyzer (SP-300 Potentiostat, Biologic). A decreasing ramp of frequency allowed us to determine the frequency dependence of the sample electrical impedance $Z^*$, which was then converted into permittivity using the following equation: $\epsilon^*(\omega)={k}/{[i\varepsilon_0\omega Z^*(\omega)]  } =\epsilon'-i\epsilon''$, where $k$ is the cell constant that has a value of $k= 0.37 \pm 0.01~\rm cm^{-1}$ calibrated in  Ref.~\onlinecite{Legrand:2023}.

\subsection{Rheometry}

Rheological measurements were performed using two complementary rotational rheometers. Recovery experiments, steady shear, and standard oscillatory measurements were carried out on a stress-controlled rheometer (MCR~302, Anton Paar). Most measurements were performed using cone-and-plate geometries (cone angle $2^\circ$), with a radius of $20$~mm and a truncation of $176~\rm \mu m$. For very soft samples with elastic moduli below 10~Pa, a larger cone-and-plate geometry (radius $25$~mm, truncation of $211~\rm \mu m$) was used to improve torque resolution, whereas for the stiffest samples (elastic moduli above 10 kPa), a parallel-plate geometry (radius 20~mm, gap 1~mm) was employed. In all cases, the stator surface was smooth, and the rotor was sandblasted to minimize wall slip.

Time-resolved mechanical spectroscopy and orthogonal superposition measurements were performed using a strain-controlled rheometer (ARES-G2, TA Instruments). Time-resolved spectra were acquired using a cone-and-plate geometry (diameter $40$~mm, cone angle $2^\circ$, truncation $46~\rm \mu m$). Orthogonal superposition experiments were conducted using a double-wall slotted Couette geometry, identical to that described in Ref.~\onlinecite{Colombo:2017}, which enables the superposition of a small-amplitude oscillatory deformation orthogonal to a steady primary shear flow. In all cases, the temperature was set to $T=22^{\circ}$C through a Peltier module. 

\subsection{Rheo-TRUSAXS measurements}

Time-resolved ultra-small-angle X-ray scattering (rheo-TRUSAXS) experiments were performed at the ID02 beamline of the ESRF (Grenoble, France) using a monochromatic X-ray beam of wavelength $1.013$~\AA~  ($12.23~\rm keV$). A stress-controlled rheometer (Haake RS6000, Thermo Scientific) equipped with a custom polycarbonate Couette cell was mounted on the beamline. The cell consisted of a rotating inner bob (radius 10~mm) and a stationary outer cup (radius 11~mm), with a sample height of 40~mm.
The X-ray beam was aligned along the velocity-gradient direction (radial configuration), so that scattering patterns were collected in the velocity-vorticity plane. A sample-to-detector distance of  $30.7~\rm m$ provided access to scattering vectors in the range $2\times 10^{-3} \leq q \leq 10^{-1}~\rm nm^{-1}$. Two-dimensional scattering patterns were recorded using an Eiger2-4M detector and processed following the standard ID02 reduction procedure, including transmission normalization and background subtraction.\cite{Narayanan:2022} 

In the explored $q$-range, the scattering signal is dominated by the CB phase due to its much larger scattering length density contrast with respect to water and CMC.\cite{Legrand:2026} To isolate the contribution of the CB microstructure, background scattering from the empty cell filled with water and from a pure CMC solution was systematically measured and subtracted, accounting for the actual CMC concentration. This rheo-TRUSAXS configuration enables time-resolved measurements of the CB network under shear, which are analyzed during shear rate cycles in Section~\ref{sec:rheoTRUSAXS}.

\section{\label{sec:aging} Aging dynamics in CB-CMC hydrogels}

\begin{figure*}[!th]
    \centering
    \includegraphics[width=\linewidth]{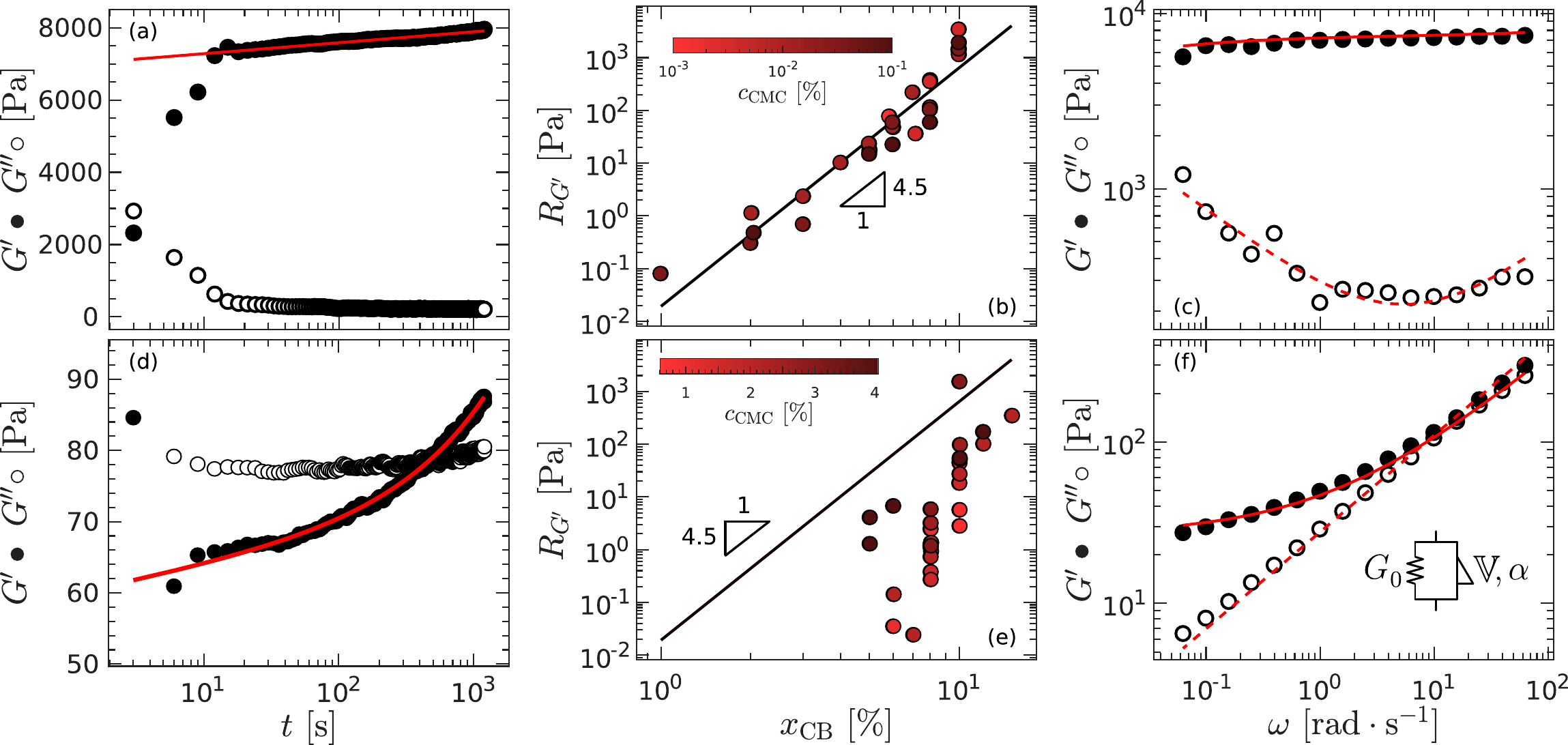}
    \caption{\label{fig:1Hz_recovery} Recovery dynamics of conductive (top row) and insulating (bottom row) CB-CMC hydrogels following a $180~\rm s$ shear step at $\dot{\gamma} = 500 ~\rm s^{-1}$, monitored by small-amplitude oscillatory shear ($\omega= 2\pi~\rm rad.s^{-1}$, $\gamma_0=$0.03\% for the conductive hydrogels and $\gamma_0=$1\% for the insulating hydrogels). 
    (a,d) Representative time evolution of the elastic modulus $G'$ and viscous modulus $G''$  for (a) a conductive hydrogel ($c_{\rm CMC} = 0.01~\%$, $x_{\rm CB} = 8 \% $) and (d) an insulating hydrogel ($c_{\rm CMC} = 3~\%$, $x_{\rm CB} = 8 \% $). The red line in (a) is the best logarithmic fit, $G'(t)=G'(t_0)+R_{G'}\log(t/t_0)$. The red curve in (d) is the best fit using the combined logarithmic and power-law recovery model [Eq.~\eqref{eq:aging}]. (b,e) Recovery rate $R_{G'}$ as a function of the carbon black content $x_{\rm CB}$ for (b) conductive and (e) insulating hydrogels. Red shades correspond to different CMC concentrations ranging from 10$^{-3}$ to 4~\%. The black line in (b) is the power-law fit: $R_{G'}\propto x_{\rm CB}^{4.5}$. The same power law is shown in (e) for comparison with the data in the insulating regime. (c,f) Representative linear viscoelastic spectra of (c) conductive ($c_{\rm CMC} = 0.01~\%$, $x_{\rm CB} = 8 \% $) and (f) insulating ($c_{\rm CMC} = 3~\%$, $x_{\rm CB} = 8 \% $) hydrogels. In (c), 
    the red curves correspond to a modified SGR model,\cite{Sollich:1997,Purnomo:2006} -- see details in Ref.~\onlinecite{Legrand:2023} In (f), the red curves show the best fit with the Fractional Kelvin-Voigt (FKV) model sketched in the inset [Eq.~\eqref{eq:FKV}].}
\end{figure*}

In this first section, we compare the aging dynamics of conductive and insulating CB-CMC hydrogels. In practice, their properties are monitored at rest, following a preshear that brings the sample into a liquid-like state. 

\subsection{Insights from oscillatory measurements at fixed frequency}
\label{sec:oscillatory}

Following a preshear at $\dot{\gamma} = 500 ~\rm s^{-1}$ applied for $180$~s, CB-CMC hydrogels display two distinct recovery dynamics depending on their composition. Below, we discuss the cases of conductive and insulating CB-CMC hydrogels separately.

\subsubsection{Conductive CB-CMC hydrogels}

Conductive CB-CMC hydrogels recover their elastic properties rapidly, with $G' \gg G''$ within a few seconds after flow cessation. This behavior is robustly observed over a broad range of compositions. 
Moreover, the elastic modulus $G'$ grows logarithmically with time for $t \gtrsim 10~\rm s$  as illustrated in Fig.~\ref{fig:1Hz_recovery}(a). The rate of growth, quantified by $R_{G'}=\mathrm{d}G'/\mathrm{d}(\log_{10} t)$ over a broad range of compositions, increases for increasing carbon black content, as shown in Fig.~\ref{fig:1Hz_recovery}(b). We find a robust power-law dependence: $R_{G'} \propto x_{\rm CB}^{4.5}$, irrespective of the CMC content, which shows that the recovery of conductive CB-CMC hydrogels is driven by the carbon black particles. Notably, this exponent coincides with that reported for the CB-content dependence of the plateau elastic modulus itself $G'_0$ measured after recovery.\cite{Legrand:2023} As a result, the relative aging rate $R_{G'}/G'_0$ -- the relative stiffening per decade of time -- is essentially independent of CB content, even though $R_{G'}$ itself spans several decades across the compositions investigated. This observation is consistent with a picture in which the network's elasticity and its logarithmic growth both originate from the same population of independently aging CB-CB contacts: increasing $x_{CB}$ increases the number of such load-bearing junctions -- and hence both $G'_0$ and $R_{G'}$ with a common exponent -- without changing the fractional rate at which each individual contact strengthens with time, consistent with a contact-level aging mechanism.

Such a slow recovery is reminiscent of the logarithmic aging reported in colloidal gels of polyethylene oxide-stabilized silica particles,\cite{Derec:2003} and precipitated silica gels.\cite{Wang:2014} It also recalls the near-logarithmic aging observed in dense silica suspensions that was linked to the increase in contact bending rigidity between particles in the absence of large-scale rearrangement.\cite{Bonacci:2020,Bonacci:2022} 
In the present case, the logarithmic growth of $G'$ is consistent with the slow mechanical relaxation of an already percolated carbon black network following flow cessation. While the connectivity of the network is rapidly re-established after preshear, local rearrangements progressively strengthen the interparticle contacts. The broad distribution of timescales associated with these activated restructuring events naturally gives rise to the observed logarithmic aging, as commonly observed in soft glassy materials.\cite{Joshi:2014} 

The fact that the aging rate $R_{G'}$ depends only on the carbon black content irrespective of the CMC concentration, shows that the rheology of conductive CB-CMC hydrogels is dominated by the CB network across the entire range of compositions investigated: increasing the amount of polymer does not measurably alter the aging response. This further supports the view that this process is governed by the carbon black network, whereas the adsorbed CMC mainly acts as a stabilizing agent and does not contribute significantly to the mechanical response.

At long times, the linear viscoelastic spectrum of conductive gels shows glassy-like features, as illustrated in Fig.~\ref{fig:1Hz_recovery}(c) for a hydrogel with $c_{\rm CMC}=0.01$\% and $x_{\rm CB}=8$\%. $G'$ is nearly frequency independent, while $G''$ shows a decreasing trend at low frequency, followed by a moderate upturn at larger frequencies. All conductive compositions display a similar spectrum, which is well described by a modified version of the Soft Glassy Rheology (SGR) model,\cite{Sollich:1997,Purnomo:2006} as extensively discussed in ref.~\onlinecite{Legrand:2023}.

\subsubsection{Insulating CB-CMC hydrogels}
\label{sec:aginginsulating}

In contrast, the recovery dynamics of insulating CB-CMC hydrogels, illustrated in Fig.~\ref{fig:1Hz_recovery}(d), progressively get faster than logarithmic and strongly depend on the CB content. Immediately after flow cessation, the material behaves predominantly as a viscoelastic liquid with $G''> G'$, which persists over much longer timescales than for conductive CB-CMC hydrogels. The subsequent crossover to an elastic-dominated response occurs earlier as the CB content is increased.\cite{Legrand:2023} Over the experimentally accessible time window, the evolution of $G'$ is well described by the sum of a logarithmic and a power-law contribution: 
\begin{equation} 
G' = G'_b+ R_{G'}\log_{10} \left(\frac{t}{t_0}\right)+A\left(\frac{t}{t_0}\right)^m, \label{eq:aging}
\end{equation}
where $R_{G'}$ denotes the logarithmic aging rate, while $A$ and $m$ characterize the amplitude and the exponent of the power-law contribution, respectively, and $G'_b$ is a baseline offset -- note that $G'_b \neq G'(t_0)$, since the power-law term contributes already at $t=t_0$. The parameter $t_0$ is arbitrarily set to $t_0=1$~s to ensure consistency in the units; any other value would solely lead to a change in $G'_b$ and $A$ values. 

The coexistence of logarithmic and power-law terms suggests that the recovery of $G'$ results from two distinct relaxation mechanisms. Figure~\ref{fig:1Hz_recovery}(e) shows that the logarithmic aging rate $R_{G'}$ tends to increase with the CB content although, unlike in the conductive regime, the data do not follow a clear power law and instead show significant scatter across the range of CMC concentrations investigated.
This scatter reflects the fact that $R_{G'}$ depends on both the polymer concentration and the CB content. This is consistent with the different microstructure of insulating CB-CMC hydrogels, in which the CMC forms the continuous load-bearing network, while the CB particles act as transient physical crosslinks. The recovery dynamics are therefore governed by the coupled evolution of both constituents rather than by the carbon black alone.
By contrast, neither the amplitude $A$ nor the exponent $m$ exhibit a systematic dependence on either the CB or the CMC content (see Fig.~\ref{Fig:paramters_TS_powerlaw} in Appendix~\ref{appendix:Fitaging}). The exponent $m$ nevertheless varies appreciably across compositions, typically ranging from approximately $0.5$ to $1.5$. 
Thus, while the power-law recovery cannot be considered composition independent, its variations do not reveal a clear trend with either constituent. This contrasts with the logarithmic contribution, whose amplitude is more directly correlated with the CB content, particularly in the conductive regime.

\begin{figure}[!t]
    \centering
    \includegraphics[width=0.95\columnwidth]{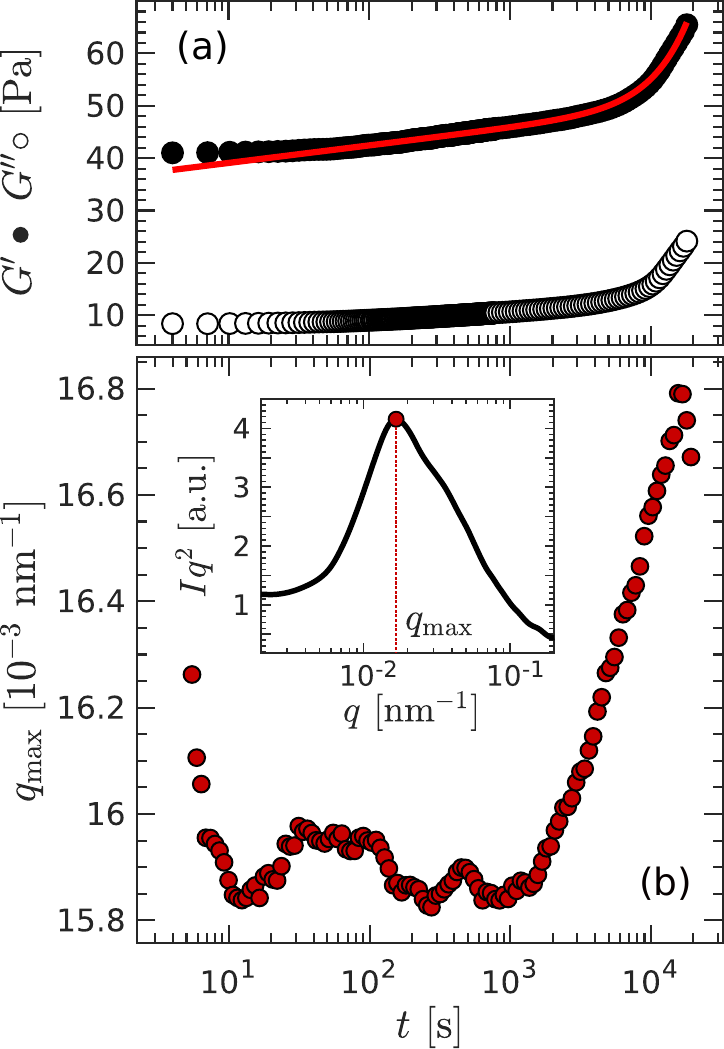}
    \caption{Rheo-USAXS measurement during the recovery of an insulating CB-CMC hydrogel ($c_{\rm CMC} = 3~\%$ and  $x_{\rm CB} = 8 \% $) following a preshear at $\dot{\gamma} = 500 ~\rm s^{-1}$ applied for $180~\rm s$. (a) Time evolution of the elastic and viscous moduli, $G'$ and $G''$, measured during recovery. The red line is the best fit with Eq.~\ref{eq:aging}, with $G'_b=35.7~\rm Pa$, $R_{G'}=3.34~\rm Pa$, $A=1.35\times 10^{-6}~\rm Pa$, $m=1.66$, and $t_0=1~\rm s$.  (b) Time evolution of the position $q_{\rm max}$ of the maximum in the Kratky representation, $Iq^2(q)$, of the USAXS intensity. The nearly constant value of $q_{\rm max}$ during the early stage ($t \lesssim 10^3~\rm s$) is followed by a slight increase at longer times, indicating a small decrease in the characteristic spacing between neighboring CB particles. Inset: Kratky representation $Iq^2(q)$ of the scattering intensity measured at time $t=5$~s; the red dot marks the maximum located at $q_{\rm max}$.
} \label{Fig:USAXSaging}
\end{figure}

This interpretation is qualitatively supported by the rheo-USAXS measurements performed during recovery (Fig.~\ref{Fig:USAXSaging}). During the early stage of the recovery ($t \lesssim 10^3~\rm s$), over which the growth of $G'$ is dominated by, though not strictly reducible to, the logarithmic term of Eq.~\eqref{eq:aging} -- the relative weight of the logarithmic and power-law contributions close to $t_0$ in fact depends on the choice of $t_0$ itself -- the position of the USAXS peak remains essentially unchanged despite the increase of the elastic modulus, suggesting that elasticity develops primarily through local rearrangements within an almost fixed CB spatial distribution. At longer times, the power-law regime coincides with a measurable increase of $q_{\rm max}$, from about $15.8\times 10^{-3}$ to $16.8\times 10^{-3}~\rm nm^{-1}$ [Fig.~\ref{Fig:USAXSaging}(b)], consistent with a small yet significant decrease in the characteristic spacing between neighboring CB particles, $d=2\pi/q_{\rm max}$, from about $400~\rm nm$ to $375~\rm nm$, i.e., by about 6\%. 
These observations suggest that the early recovery is primarily associated with the reconfiguration of the CMC network around a nearly fixed CB scaffold. Note that this is not at odds with the logarithmic aging rate being governed by the density of load-bearing CB junctions. Indeed, the number and strength of these junctions can increase as CMC chains progressively engage with a given, essentially static, population of CB particles, without requiring any large-scale displacement of the CB particles. The spatial organization of the CB particles, in this sense, remains nearly unchanged during this stage. At longer times ($t \gtrsim 10^3~\rm s$), the stronger power-law stiffening is accompanied by a measurable change in the  CB spatial correlations, suggesting that structural rearrangements of the physical crosslinkers increasingly contribute to the mechanical recovery.

Finally, the linear viscoelastic spectrum of insulating CB-CMC hydrogels is well described by a Fractional Kelvin-Voigt solid model, hereafter denoted FKV, as illustrated in Fig.~\ref{fig:1Hz_recovery}(f) and extensively discussed in Ref.~\onlinecite{Legrand:2023}. This mechanical model is a particular case of the generalized fractional Kelvin-Voigt model and consists of an elastic spring of modulus $G_0$, accounting for the low-frequency elastic plateau, connected in parallel with a spring-pot. The latter element is intermediate between an elastic spring and a dashpot, and characterized by the following constitutive equation $\sigma = \mathbb{V}\mathrm{d}^\alpha\gamma/\mathrm{d} t^\alpha$, where $\alpha$ is a dimensionless exponent, and $\mathbb{V}$ is a quasi-property with dimensions $[\rm Pa.s^\alpha]$.\cite{Jaishankar:2013,Bonfanti:2020} The FKV model corresponds to the following complex modulus: 
\begin{equation}
G^*=G_0+\mathbb{V}(i\omega)^\alpha, \label{eq:FKV} 
\end{equation} 
where the exponent $\alpha$ captures the power-law frequency dependence of the viscoelastic response. As previously shown in Ref.~\onlinecite{Legrand:2023}, the FKV model accurately describes the linear viscoelastic spectra of insulating gels over a broad range of compositions, which collapse onto a single master curve after appropriate rescaling, highlighting a time-concentration superposition principle, which is not further discussed here.

\begin{figure*}[!th]
    \centering
    \includegraphics[width=\linewidth]{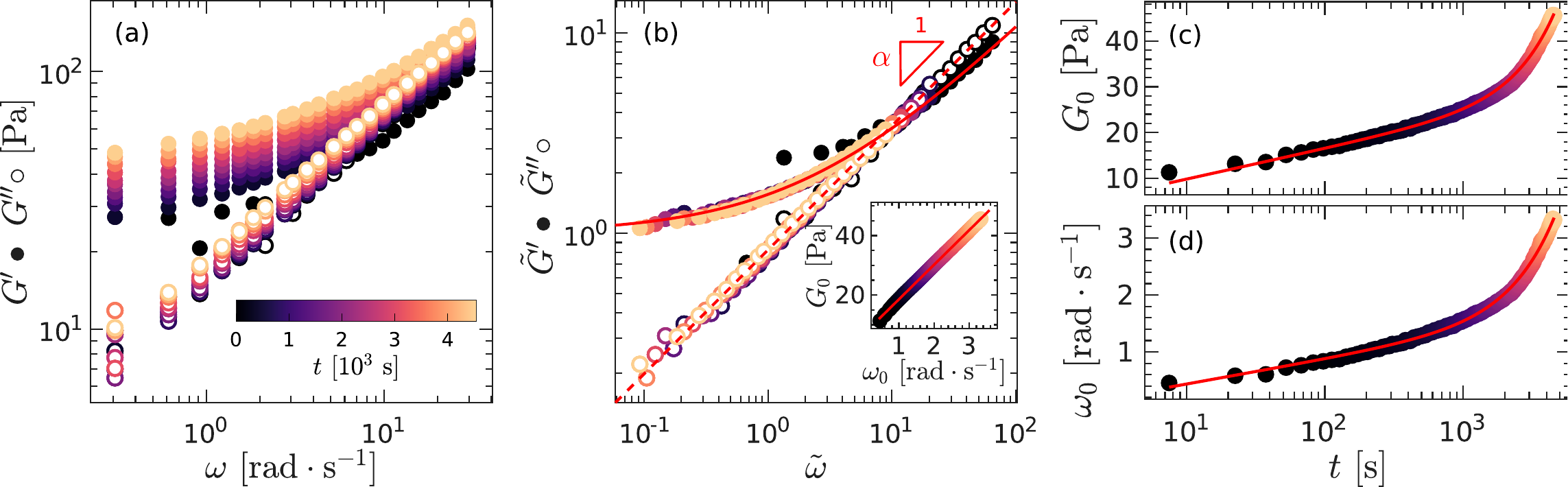}
    \caption{Time-resolved mechanical spectroscopy during the recovery of an insulating CB-CMC hydrogel ($c_{\rm CMC} = 3~\%$ and  $x_{\rm CB} = 8 \% $) following a preshear at $\dot{\gamma} = 500 ~\rm s^{-1}$ applied for $180~\rm s$. (a) Evolution of the viscoelastic spectra $G'(\omega)$ and $G''(\omega)$ measured using an OWCh sequence\cite{Geri:2018} over the frequency range $\omega= 0.3-30~\rm rad.s^{-1}$. Colors indicate the time elapsed since the end of the preshear. (b) Collapse of the viscoelastic spectra onto a single master curve after rescaling the viscoelastic moduli and the frequency according to $\tilde G'=G'/G_0$, $\tilde G''=G''/G_0$, and $\tilde \omega = \omega/\omega_0$ where $G_0$ and $\omega_0=(G_0/\mathbb{V})^{1/\alpha}$ are obtained by fitting each spectrum with the Fractional Kelvin-Voigt (FKV) model [see Eq.~(\ref{eq:FKV})] using a fixed exponent $\alpha = 0.62$. The red curves correspond to the normalized FKV prediction, $\tilde{G}^\star = 1 + \left( i \tilde{\omega} \right)^\alpha$. Inset: $G_0$ vs.~$\omega_0$; the red line is the best linear fit of the data. 
    For clarity, only one third of the frequency points in each spectrum and one third of the measured spectra are displayed in (a) and (b). (c,d) Time evolution of the fitted parameters (c) $G_0$ and (d) $\omega_0$, where the red lines correspond to a fit following Eq.~\eqref{eq:aging} leading to $m=1.44$ for $G_0$ and $m=1.33$ for $\omega_0$.
    }
    \label{fig:chirp_recovery}
\end{figure*}

\subsection{Insights from time-resolved mechanical spectroscopy}
\label{sec:TRMS}

To better characterize the time-dependent behavior of insulating CB-CMC hydrogels, we perform time-resolved mechanical spectroscopy and monitor the evolution of the viscoelastic spectrum immediately after flow cessation. Note that we restrict this time-resolved analysis to insulating hydrogels because their recovery spans several decades in time (from a few seconds to $\sim 10^4\rm s$, Sec.~\ref{sec:aginginsulating}), whereas conductive hydrogels reach an essentially glassy, quasi-stationary spectrum within a few seconds of flow cessation [Fig.~\ref{fig:1Hz_recovery}(a,c)] that is already well-captured by the modified SGR model of Ref.~\onlinecite{Legrand:2023} without requiring a fully time-resolved treatment. 

Optimally Windowed Chirp signals\cite{Geri:2018} (duration $15$~s, repeated 300 times) are used to record the viscoelastic spectrum of an insulating CB-CMC hydrogel ($c_{\rm CMC} = 3\%$ and $x_{\rm CB} = 8\%$) in a time-resolved manner over a total duration of 4500~s. The measurements span two decades in frequency, from $\omega= 0.3$ to $30~$rad.s$^{-1}$. The raw spectra are shown in Fig.~\ref{fig:chirp_recovery}(a). Remarkably, all viscoelastic spectra display a similar shape, as early as a few seconds after flow cessation. From the very first measurement, the sample exhibits a solid-like response, dominated by the elastic component at low frequencies. 
All spectra are well described by the FKV model --see Eq.~\eqref{eq:FKV}-- with a fixed exponent $\alpha =0.62$. The two remaining fitting parameters, $G_0$ and $\mathbb{V}$, are reported in Fig.~\ref{fig:chirp_recovery}(c)-(d) in the form of two characteristic dimensional quantities: an elasticity scale $G_0$ and a characteristic frequency scale $\omega_0=(G_0/\mathbb{V})^{1/\alpha}$. Both parameters increase with time, and are well described by Eq.~\ref{eq:aging}, i.e., a combination of a logarithmic growth 
up to approximately $t \simeq 2000~\rm s$, beyond which they exhibit an upturn compatible with a steeper power-law increase: $G_0\sim t^{n_G}$ and $\omega_0\sim t^{n_{\omega}}$, with $n_G=1.44$ and $n_{\omega}=1.33$. This crossover time is comparable to the one identified independently from the $G'(t)$ recovery curve, where the power-law term starts to dominate over the logarithmic one, supporting a consistent picture in which two successive relaxation regimes govern the build-up of elasticity in insulating CB-CMC hydrogels.

In addition, these two parameters $G_0$ and $\omega_0$ can further be used to rescale all viscoelastic spectra by plotting the normalized moduli $\tilde G'=G'/G_0$ and $\tilde G''=G''/G_0$ as a function of the reduced frequency $\tilde \omega = \omega/\omega_0$. As shown in Fig.~\ref{fig:chirp_recovery}(b), all viscoelastic spectra collapse onto a single master curve. This collapse demonstrates that insulating CB-CMC hydrogels obey a time-aging (or time-connectivity) superposition principle.\cite{Keshavarz:2021,Joshi:2025} Interestingly, the fact that a single time-connectivity superposition captures both the logarithmic and power-law regimes of Sec.~\ref{sec:oscillatory} indicates that the qualitative nature of the interactions between the gel constituents does not change over the course of aging: CB particles (or clusters of particles) remain isolated, non-percolating physical crosslinkers throughout, and the two successive regimes should therefore be understood as reflecting an evolving functionality (connectivity) of the load-bearing nodes rather than a change in the nature of the interactions themselves.

\begin{figure*}[!th]
    \centering
    \includegraphics[width=\linewidth]{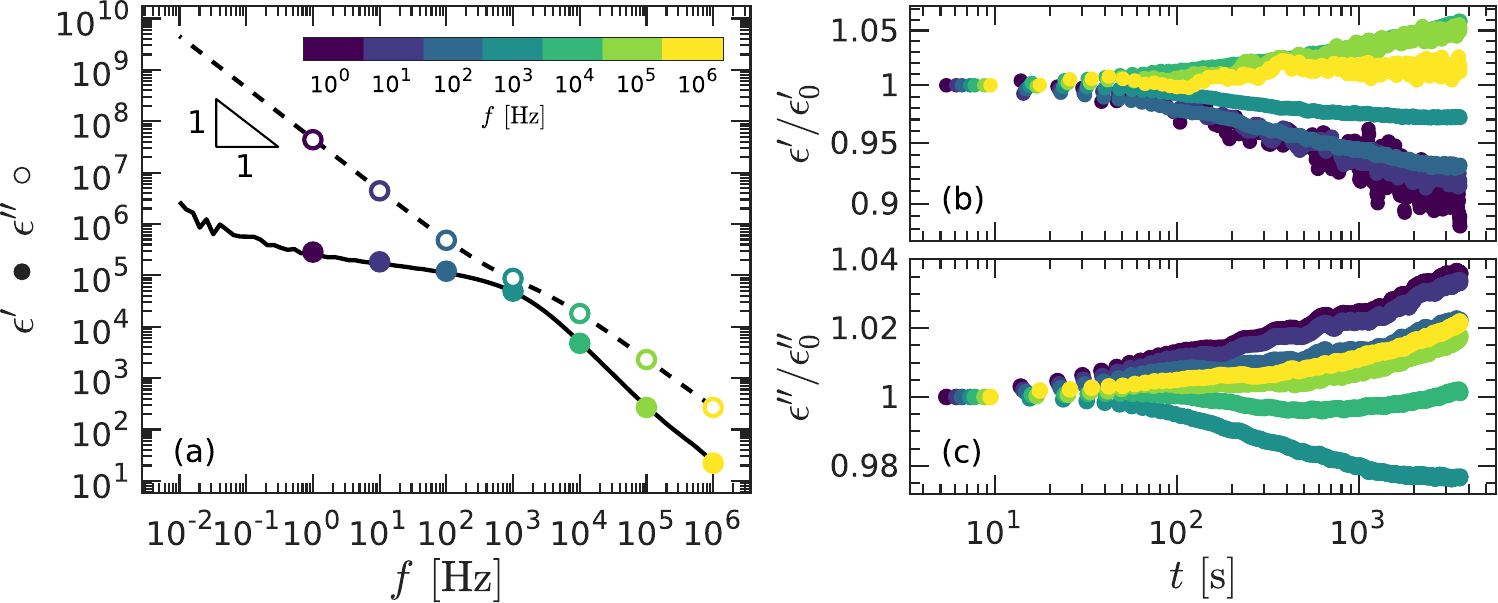}
    \caption{Time evolution of the complex relative permittivity $\epsilon^\star = \epsilon^\p - i \epsilon^\pp$ of a conductive CB-CMC hydrogel ($c_{\rm CMC} = 0.01\%$ and  $x_{\rm CB} = 8 \% $) as a function of frequency. (a) Frequency-dependent permittivity spectra measured immediately after loading the sample into the measurement cell (circles) and after approximately 1~h (black lines). The solid and dashed lines correspond to $\epsilon'$ and $\epsilon''$, respectively; the final spectrum was measured over an extended frequency range.
    (b, c) Time evolution of the normalized permittivity components (b)~$\epsilon'/\epsilon'_0$ and (c)~$\epsilon''/\epsilon''_0$, measured at selected frequencies, with $\epsilon'_0$ and $\epsilon''_0$ the first measured values of $\epsilon'$ and $\epsilon''$. Colors indicate the measurement frequency ($1~\rm Hz$ to $1~\rm MHz$, one frequency per decade), as shown by the color bar in (a). The triangle in (a) highlights the low-frequency power-law dependence $\epsilon'' \propto f^{-1}$.}
    \label{fig:permittivity_conductive}
\end{figure*}

\subsection{Insights from Electrical Impedance Spectroscopy}

\begin{figure*}[!th]
    \centering
    \includegraphics[width=\linewidth]{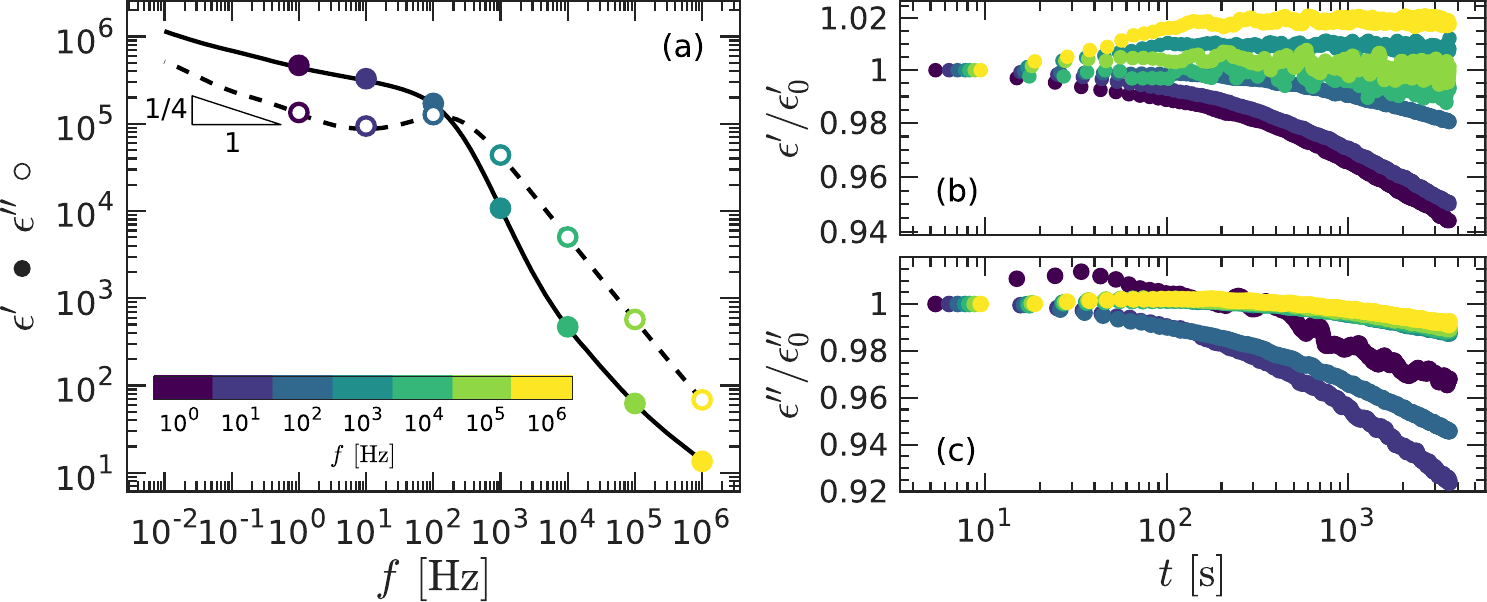}
    \caption{Time evolution of the complex relative permittivity $\epsilon^\star = \epsilon^\p - i \epsilon^\pp$ of an insulating CB-CMC hydrogel ($c_{\rm CMC} = 2\%$ and  $x_{\rm CB} = 8 \% $). (a) Relative permittivity measured immediately after loading the sample into the measurement cell (circles) and after approximately 1~h (black lines). The solid and dashed lines correspond to $\epsilon'$ and $\epsilon''$, respectively; the final spectrum was measured over an extended frequency range. (b,c) Time evolution of the normalized permittivity components (b)~$\epsilon'/\epsilon'_0$ and (c)~$\epsilon''/\epsilon''_0$ measured at selected frequencies, with $\epsilon'_0$ and $\epsilon''_0$ the first measured values of $\epsilon'$ and $\epsilon''$. Colors indicate the measurement frequency ($1~\rm Hz$ to $1~\rm MHz$, one frequency per decade), as shown by the color bar in (a). The triangle in (a) highlights the low-frequency power-law dependence $\epsilon'' \propto f^{-1/4}$.}
    \label{fig:permittivity_insulating}
\end{figure*}

To complement mechanical measurements, we have performed electrical impedance spectroscopy on both conductive and insulating CB-CMC hydrogels. Experiments were performed in a static cell, which does not allow any preshear.
Because these measurements start from the sample as-loaded state rather than from a controlled, sheared liquid-like reference state (contrary to the mechanical recovery protocol of Sec.~\ref{sec:oscillatory}, which follows a large-amplitude preshear), the resulting temporal changes in $\epsilon'$ and $\epsilon''$ reported below are comparatively modest and should not be compared quantitatively, in amplitude, to the aging rates or superposition principles established from rheology.
In practice, the samples were loaded, and the complex impedance $Z^*$ was recorded every $10$~s for an hour at several frequencies from $1$~Hz to $1$~MHz.
Finally, a terminal spectrum was recorded after an hour over an extended range of frequencies, from $1$~MHz down to $10^{-2}$~Hz. The results are shown in Figs.~\ref{fig:permittivity_conductive} and \ref{fig:permittivity_insulating} for the conductive and insulating hydrogels, respectively.

\subsubsection{Electric properties of conductive CB-CMC hydrogels}

Figure~\ref{fig:permittivity_conductive} presents the real and imaginary parts of the complex dielectric permittivity, $\epsilon'$ and $\epsilon''$, respectively, of a conductive CB-CMC hydrogel ($c_{\rm CMC} = 0.01\%$ and  $x_{\rm CB} = 8 \% $). Fig.~\ref{fig:permittivity_conductive}(a) shows the frequency dependence of $\epsilon'$ and $\epsilon''$. Colored symbols at discrete frequencies were recorded over an hour, starting shortly after loading the sample into the measurement cell, while the black lines were recorded after one hour at rest over an extended frequency range. Figs.~\ref{fig:permittivity_conductive}(b)-(c) report the temporal evolution of $\epsilon'$ and $\epsilon''$ at the selected frequencies highlighted in Fig.~\ref{fig:permittivity_conductive}(a).

At the beginning of the measurement, the electric spectrum already displays the characteristic features of a percolated conductive system.\cite{Meier:2008,Azizi:2019} In the low-frequency regime, the imaginary part follows a clear power-law decay,  $\epsilon'' \propto f^{-1}$, which is the signature of continuous electronic DC transport through the percolated carbon black network.\cite{Kremer:2002} 
Concurrently, the high magnitude of $\epsilon'$ at low frequencies together with the visible shoulder of $\epsilon''$ at $f=10^3\ \rm Hz$ stems from Maxwell-Wagner-Sillars interfacial polarization, which arises from charge accumulation at the boundaries between the electronically conductive CB clusters and the surrounding ionically conductive aqueous medium. \cite{Sillars:1937,Kremer:2002}

Over the course of one hour, the electric response evolves modestly but systematically, with clear frequency-dependent trends. At high frequencies ($f \gg 1~\rm kHz$), the real part of the permittivity $\epsilon'$ increases slightly with time, whereas at low frequencies ($f \lesssim 1~\rm kHz$), $\epsilon'$ decreases by up to $10\%$. These opposite trends indicate that quiescent structural relaxation affects long-range charge transport and localized charge storage in contrasting ways. 

The decrease of $\epsilon'$ at low frequencies suggests a gradual decrease of large-scale polarization processes associated with the long-range charge migration, as the conductive CB network and the surrounding aqueous phase become increasingly constrained at rest. In contrast, the increase in $\epsilon'$ at high frequencies points to a local microstructural reorganization of the percolated network following the mechanical disturbance induced by sample loading; as the strained aggregate network undergoes quiescent relaxation, the subtle reorganization of local particle contacts and narrowing of sub-micron inter-particle gaps enhance the high-frequency localized polarization and charge-storage capacity of the gel architecture.\cite{Meier:2008}

The imaginary part $\epsilon''$ weakly increases with time over most of the explored frequency range, indicating a progressive enhancement of dissipative electric processes. This increase is consistent with an improved microscopic connectivity within the CB network, together with enhanced localized charge transport at CB-CB contacts through hopping or tunneling mechanisms.\cite{Medalia:1986,Balberg:1987,Karasek:1996}

Although the amplitude of the electrical aging remains modest, its systematic and frequency-dependent nature mirrors the mechanical aging observed in conductive CB-CMC hydrogels. Both responses point to a common structural origin rooted in the slow relaxation of a percolated carbon black network stabilized by the polymer, which progressively suppresses large-scale dynamics while enhancing local connectivity.

\subsubsection{Electric properties of insulating CB-CMC hydrogels}

Figure~\ref{fig:permittivity_insulating} displays the complex dielectric permittivity $\epsilon^*=\epsilon'-i\epsilon''$ of an insulating CB-CMC hydrogel ($c_{\rm CMC} = 2\%$ and  $x_{\rm CB} = 8 \% $) using the same protocol as in Fig.~\ref{fig:permittivity_conductive}. Here, for $f<100~\rm Hz$,
$\epsilon''$ is several orders of magnitude smaller than for the previous conductive CB-CMC hydrogel, which is consistent with the absence of a percolated carbon black network.\cite{Meier:2008} The electric response is therefore dominated by polarization within the CMC matrix rather than by electronic transport through carbon black. At low frequencies ($f\lesssim 100~\rm Hz$) the real part $\epsilon'$ exceeds the imaginary part $\epsilon''$, indicating that polarization processes dominate over dissipative losses in this regime. Over this frequency range, $\epsilon''$ decreases with increasing frequency following a weak power law, $[\epsilon''(f) \sim f^{-1/4}]$, markedly different from the $f^{-1}$ scaling observed in conductive samples. 
We note that this $-1/4$ exponent is not, in itself, the signature of a specific dissipative mechanism: it more likely corresponds to the descending flank, over a limited frequency window, of a single dielectric relaxation process -- most plausibly of Maxwell-Wagner-Sillars type. Nonetheless, 
this weak frequency dependence indicates that dielectric losses are no longer governed by simple DC conduction but rather by dispersive ionic dynamics and interfacial polarization processes distributed over a broad range of relaxation times, as commonly observed in polymer electrolytes and disordered ionic conductors.\cite{Chan:2020}

\begin{figure*}[!th]
    \centering
    \includegraphics[width=0.95\linewidth]{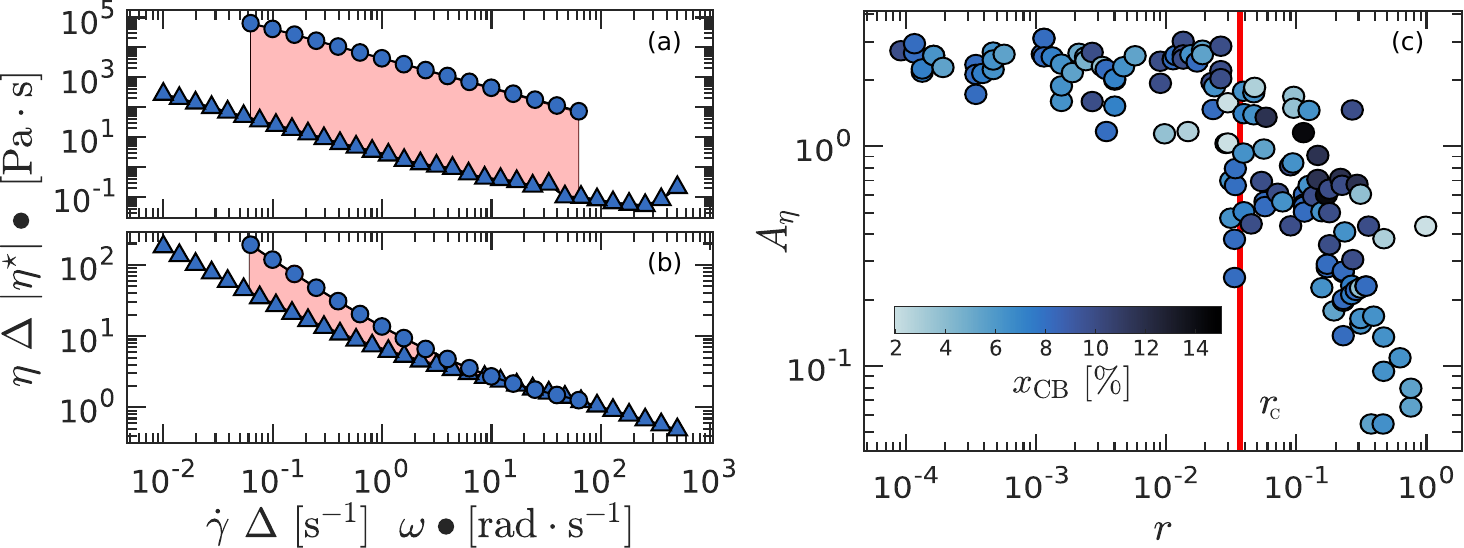}
    \caption{Assessment of the Cox-Merz empirical rule for CB-CMC hydrogels of different compositions. (a,b) Comparison between the steady-shear viscosity $\eta(\gpt)$ measured during a decreasing shear rate sweep (triangles) and complex viscosity $|\eta^\star(\omega)| = |G^\star|/ \omega $ obtained from a frequency sweep (circles). (a) Conductive CB-CMC hydrogel ($c_{\rm CMC}=$0.01\% and $x_{\rm CB}=$8\%). (b) Insulating CB-CMC hydrogel ($c_{\rm CMC}=$2\% and $x_{\rm CB}=$8\%). (c) Normalized area $A_\eta$ enclosed between $\eta(\gpt)$ and $|\eta^\star(\omega)|$ as a function of the CMC-to-CB mass ratio $r=m_{\rm CMC}/m_{\rm CB}$. The area, displayed in red in (a,b), is computed over the largest overlapping range of frequencies for which $\omega=\gpt$. Smaller values of $A_\eta$ indicate better agreement with the Cox-Merz rule. The vertical red line in (c) indicates the critical CMC-to-CB ratio $r=r_c=0.037$ separating conductive and insulating hydrogels, as previously identified by impedance spectroscopy.\cite{Legrand:2023} }
    \label{fig:cox_merz}
\end{figure*}

For $f \gtrsim 100~\rm Hz$, the trend is inverted and $\epsilon''$ becomes larger than $\epsilon'$, reflecting the prominent role of dissipative processes at shorter timescales. This high-frequency regime is characterized by a massive drop in $\epsilon'$ by nearly four orders of magnitude. The sharp decrease of both components and the fact that $\epsilon''>\epsilon'$ represent the characteristic high-frequency tail of a localized dielectric relaxation process. This feature is consistent with restricted, short-range ion hopping -- potentially involving the exchange of mobile cations between free and CMC-bound (carboxylate) state -- and localized polarization dynamics governed by a distinct relaxation timescale within the constrained polymer-colloid matrix, analogous to the intermediate-frequency relaxation of free counterions confined within the correlation length of a polyelectrolyte network, and to the bound/free counterion exchange mechanism proposed for polyelectrolyte solutions more generally.\cite{Bordi:2004,Choi:2023}

The time evolution at fixed frequency reported in Figs.~\ref{fig:permittivity_insulating}(b)-(c) further highlights the insulating nature of these hydrogels.
At high frequencies, $\epsilon'$ remains essentially constant over the duration of the experiment, whereas at low frequencies, $\epsilon'$ decreases by up to about 6\% over one hour, reflecting a progressive confinement and partial immobilization of slow ionic polarization modes. As the polymer-CB network stiffens, ions become increasingly constrained, reducing their ability to reorganize coherently over large distances and long times, which leads to a diminished low-frequency polarizability.
The decrease of $\epsilon''$ with time at low frequency can be related to the concurrent rheological aging of the sample: as the elastic modulus grows via the power-law contribution identified in Sec.~\ref{sec:aginginsulating}, the effective viscosity experienced by the mobile ionic charge carriers is expected to increase correspondingly, lowering their diffusion coefficient and hence the DC-like conductivity contribution to $\epsilon''$ at low frequency.

Taken together, the rheological and electrical measurements reveal that conductive and insulating CB-CMC hydrogels recover through fundamentally different mechanisms, reflecting their distinct underlying microstructures. While the previous section focused on the evolution of these systems at rest following flow cessation, the following section addresses the complementary question of how these microstructures respond under continuous shear.

\section{\label{sec:flow} Continuous shear of CB-CMC hydrogels}

In this section, we investigate the rheology of CB-CMC hydrogels under continuous shear. We first compare the linear and nonlinear rheological responses by assessing the validity of the Cox-Merz empirical rule over a broad range of hydrogel compositions (Sec.~\ref{sec:CoxMerz}). Remarkably, the validity of the Cox-Merz rule changes abruptly across the critical CMC-to-CB mass ratio $r_c$ separating conductive and insulating hydrogels, providing an independent rheological signature of the transition between the two microstructural regimes. We then focus on insulating CB-CMC hydrogels, using orthogonal superposition rheometry to establish a generalized time-shear-rate superposition principle (Sec.~\ref{sec:InsulatingOSP}), before examining the structural origin of its breakdown through Rheo-TRUSAXS (Sec.~\ref{sec:rheoTRUSAXS}).  

\subsection{Cox-Merz empirical rule}
\label{sec:CoxMerz}

\begin{figure*}[!th]
    \centering
    \includegraphics[width=0.95\linewidth]{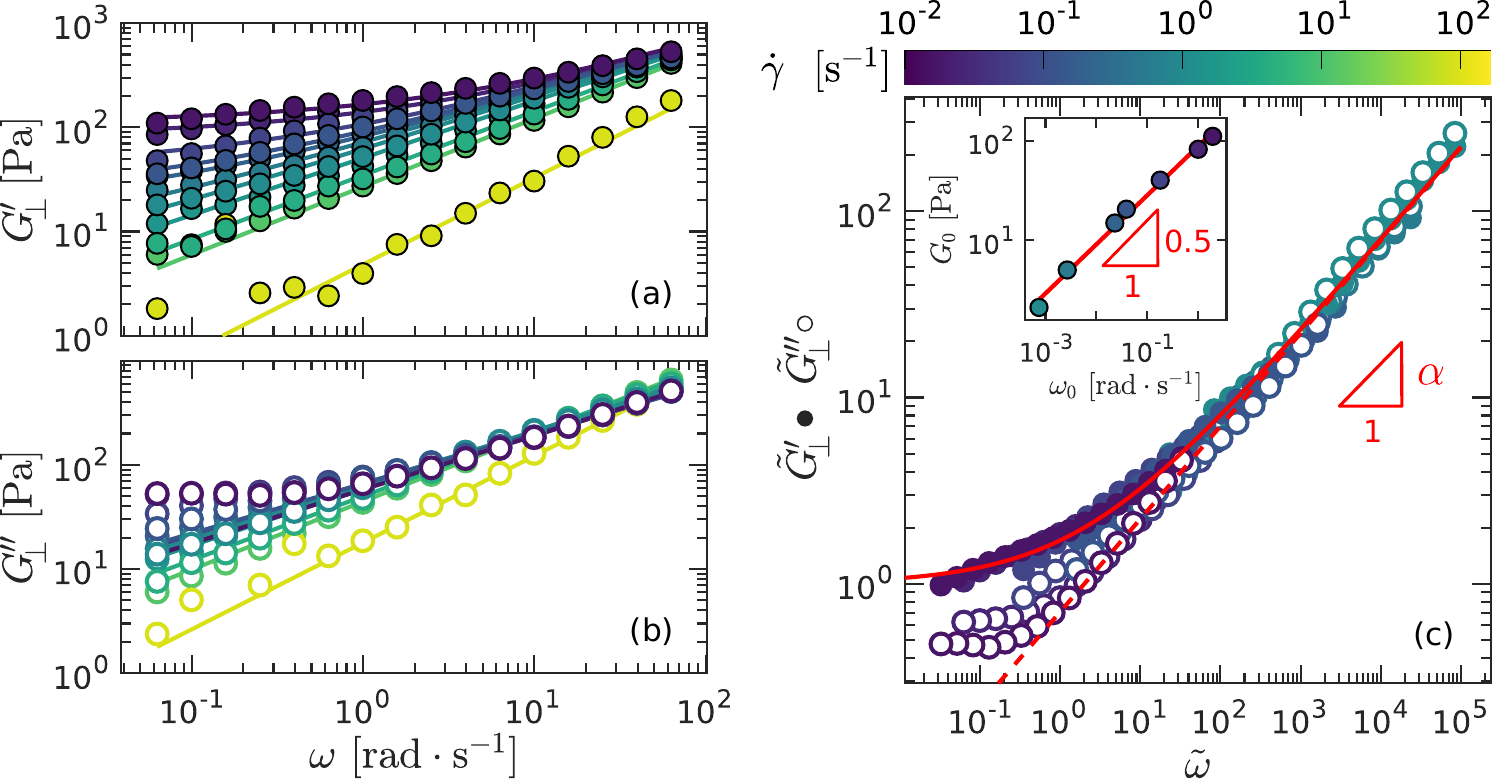}
    \caption{Orthogonal superposition rheology of an insulating CB-CMC hydrogel  ($c_{\rm CMC} = 2\%$ and $x_{\rm CB} = 10\%$). (a,b) Perpendicular elastic modulus $G'_\perp$ and viscous modulus $G''_\perp$ measured by orthogonal superposition rheometry at different imposed shear rates $\gpt$. Solid lines show fits to the Fractional Kelvin-Voigt (FKV) model [see Eq.~\eqref{eq:FKV}]. Colors indicate the imposed shear rate, with lighter shades corresponding to larger values of $\gpt$ (see color bar). (c) Collapse of the viscoelastic spectra onto a single master curve for shear rates $\gpt < \gpt^\star = 1~\rm s^{-1}$ after rescaling the moduli by $G_0$ and the frequency by $\omega_0$, both obtained from the FKV fits (see parameters in Fig.~\ref{fig:OSP_KVF}), demonstrating that a time-shear-rate superposition analogous to the time-aging superposition observed during recovery in Fig.~\ref{fig:chirp_recovery} holds at low shear rates. The red curves correspond to the normalized FKV model with $\alpha = 0.5$, where $\alpha$ is the average exponent obtained from individual fits. Inset: correlation between $G_0$ and $\omega_0$; the red line is the best power-law fit yielding $G_0 \propto \omega_0^{0.5}$.}
    \label{fig:OSP_master_curve}
\end{figure*}

Before investigating the viscoelastic response under steady shear by orthogonal superposition, we first compare the rheology measured under continuous and oscillatory shear. This comparison provides a simple way to assess whether the relationship between linear and nonlinear rheology depends on the underlying gel microstructure.
In this context, continuous shear and oscillatory rheology are, in some cases, directly comparable through the so-called Cox–Merz rule,\cite{cox1959rheology,Doraiswamy:1991,Shim:2024} which compares the steady-shear viscosity, $\eta(\dot \gamma)$, to the complex viscosity, $|\eta^*(\omega)|$, measured under oscillatory shear. The empirical rule predicts that the two quantities coincide when compared at an identical timescale, i.e., for $\omega=\dot \gamma$. A striking difference emerges between the two classes of hydrogels.
The Cox-Merz rule systematically fails for conductive samples [Fig.~\ref{fig:cox_merz}(a)], whereas it is closely satisfied for insulating CB-CMC hydrogels [Fig.~\ref{fig:cox_merz}(b)]. 

To quantify deviations from the Cox-Merz rule, we compute the normalized area $A_\eta$ enclosed between $\eta(\dot \gamma)$ and $|\eta^*(\omega)|$ over the range of timescales that is common to both measurements: 
\begin{equation}
    A_\eta=  \dfrac{1}{\log_{10}   ( \gpt_{\rm max} / \gpt_{\rm min} )} \int_{ \gpt_{\rm min}}^{ \gpt_{\rm max}} \left| \log_{10} \left(\dfrac{|\eta^\star(\omega = \gpt )|}{\eta(\gpt)} \right) \right| \mathrm{d} \log_{10} \gpt \label{eq:mat_meth_area_cox_merz}
\end{equation}
Here, $\gpt_{\rm min}$ and $\gpt_{\rm max}$ define the largest range over which the condition $\gpt=\omega$ can be satisfied, corresponding to the overlap between the frequency window explored in oscillatory shear and the shear-rate window investigated under steady shear. The area $A_\eta$ is computed numerically after linearly interpolating the $|\eta^\star(\omega)|$ and $\eta(\gpt)$ data so that both quantities are defined on identical abscissas. 

Based on more than one hundred compositions --same samples as those studied in Ref.~\onlinecite{Legrand:2026}-- $A_\eta$ remains large and nearly constant throughout the conductive regime, up to the depercolation threshold of the carbon black particles indicated by the vertical red line in Fig.~\ref{fig:cox_merz}(c). Beyond this critical ratio, where the hydrogels become electrically insulating,\cite{Legrand:2023} $A_\eta$ sharply decreases, indicating a progressively closer agreement between steady-shear and oscillatory viscosities.

This sharp transition demonstrates that the validity of the Cox-Merz rule is governed by the underlying gel microstructure.
In conductive hydrogels, the steady and complex viscosities exhibit similar dependencies on shear rate and frequency, respectively, but remain separated by an almost constant offset over the accessible range of frequencies. Such a scale-preserving deviation is distinct from the more commonly reported forms of Cox-Merz breakdown, which typically emerge at large shear rates or frequencies, for example in systems involving shear-induced disruption of transient interactions in polymer solutions,\cite{Kulicke:1980} or polymer suspensions with hard fillers, where steady and oscillatory shear can probe different dynamically evolving microstructures.\cite{Matsumoto:1975,Kitano:1980}
In contrast, insulating hydrogels satisfy the Cox-Merz empirical rule to a much better extent, although noticeable deviations remain at the lowest shear rates and frequencies, suggesting a partial decoupling between linear and nonlinear responses in this regime.
The progressive improvement of the Cox-Merz agreement with increasing CMC-to-CB ratio further supports the conclusion that departures from the rule originate from the CB network rather than from the polymer matrix.

Having established that the viscosity of insulating hydrogels remains similar under oscillatory and steady shear, we now investigate whether this correspondence extends to the frequency-dependent viscoelastic response measured under flow using orthogonal superposition rheometry.

\subsection{Insulating CB-CMC hydrogels: orthogonal superposition}
\label{sec:InsulatingOSP}

The central question addressed in this section is how the linear viscoelastic response of insulating CB-CMC hydrogels is modified when the material is subjected to a steady shear flow. In particular, we aim to determine whether the viscoelastic spectra measured \textit{under flow} and perpendicularly to the direction of shear differ qualitatively from those measured at rest, and how these changes depend on the applied shear rate. To this end, we resort to orthogonal superposition (OSP) rheometry, which provides direct access to the viscoelastic response in a direction perpendicular to the imposed flow.

In OSP experiments, a small-amplitude oscillatory deformation is applied along the vorticity direction to probe the viscoelastic spectrum $G^*_\perp(\omega)$, while a constant shear rate $\gpt$ is applied.\cite{Colombo:2017} 
Figure~\ref{fig:OSP_master_curve}(a)-(b) displays the perpendicular elastic and viscous moduli $G'_\perp$ and $G''_\perp$, measured for an insulating CB-CMC hydrogel ($c_{\rm CMC} = 2\%$ and $x_{\rm CB} = 10\%$) over shear rates ranging from $\gpt =0.01~\rm s^{-1}$ to $100~\rm s^{-1}$. 
At high frequencies, shear only weakly affects the perpendicular viscoelastic properties, particularly for $G''_\perp$, indicating that the fast, polymer-solvent-dominated dynamics are comparatively less perturbed by shear over most of the explored range -- although the slopes of $G'_\perp(\omega)$ and $G''_\perp(\omega)$ are not strictly identical across shear rates, and the highest rate investigated ($\dot \gamma=100~\rm s^{-1})$ departs noticeably from the others. In contrast, at low frequencies, both $G'_\perp$ and $G''_\perp$ are strongly affected by shear and decrease by about two decades at the lowest probed frequency as $\gpt$ increases from $10^{-2}$~s$^{-1}$ to $10^{2}$~s$^{-1}$. The low-frequency region of the spectrum is associated with the polymer network crosslinked by CB, whose connectivity is weakened under shear. 

Despite these changes, the perpendicular spectra remain remarkably well described by a Fractional Kelvin-Voigt (FKV) model over a broad range of shear rates, very similar to the viscoelastic response measured along the preshear direction both during the recovery (see Fig.~\ref{fig:chirp_recovery}) and at rest for various sample compositions.\cite{Legrand:2023} This observation suggests that, under shear, the system retains the same basic microstructural architecture --namely a CMC polymer network in which CB particles (or clusters) act as physical crosslinks-- while the efficiency of these crosslinks is progressively reduced. Crucially, this description holds only below a critical shear rate $\gpt^\star \simeq 1~\rm s^{-1}$. For $\gpt > \gpt^\star$, the elastic spring element vanishes ($G_0 \rightarrow 0$), and the FKV model reduces to a single spring-pot element, characterized by the absence of a well-defined modulus or frequency scale. For shear rates $\gpt < \gpt^\star$, both a characteristic modulus $G_0$ and a characteristic frequency $\omega_0$ are defined, enabling a time–shear-rate superposition of the perpendicular spectra. 
As shown in Fig.~\ref{fig:OSP_master_curve}(c), all perpendicular viscoelastic spectra collapse onto a master curve spanning nearly six decades in rescaled frequency, although a systematic deviation from the Fractional Kelvin-Voigt prediction is visible in $\tilde G''_{\perp}$ at the lowest rescaled frequencies. This deviation may point to an additional, slower relaxation process not captured by the two-parameter FKV description, and would deserve further investigation. Strikingly, the resulting scaling quantitatively differs from the one observed at rest: here, the characteristic modulus follows $G_0 \sim \omega_0 ^{0.5}$, whereas during recovery over time [see inset in Fig.~\ref{fig:chirp_recovery}(b)] as well as at rest across samples of various compositions,\cite{Legrand:2023} one finds $G_0 \sim \omega_0$. 
This sharp difference in scaling points to a shear-induced modification of the network connectivity, rather than a mere shift of relaxation times. Moreover, $\tilde G'_{\perp}$ and $\tilde G''_{\perp}$ become nearly superimposed over the rescaled master curve [Fig.~\ref{fig:OSP_master_curve}(c)], in contrast with the well-separated moduli found at rest [Fig.~\ref{fig:chirp_recovery}(b)] -- a point we return to in Sec.~\ref{sec:discussion}.

\begin{figure}[!t]
    \centering
    \includegraphics[width=0.95\linewidth]{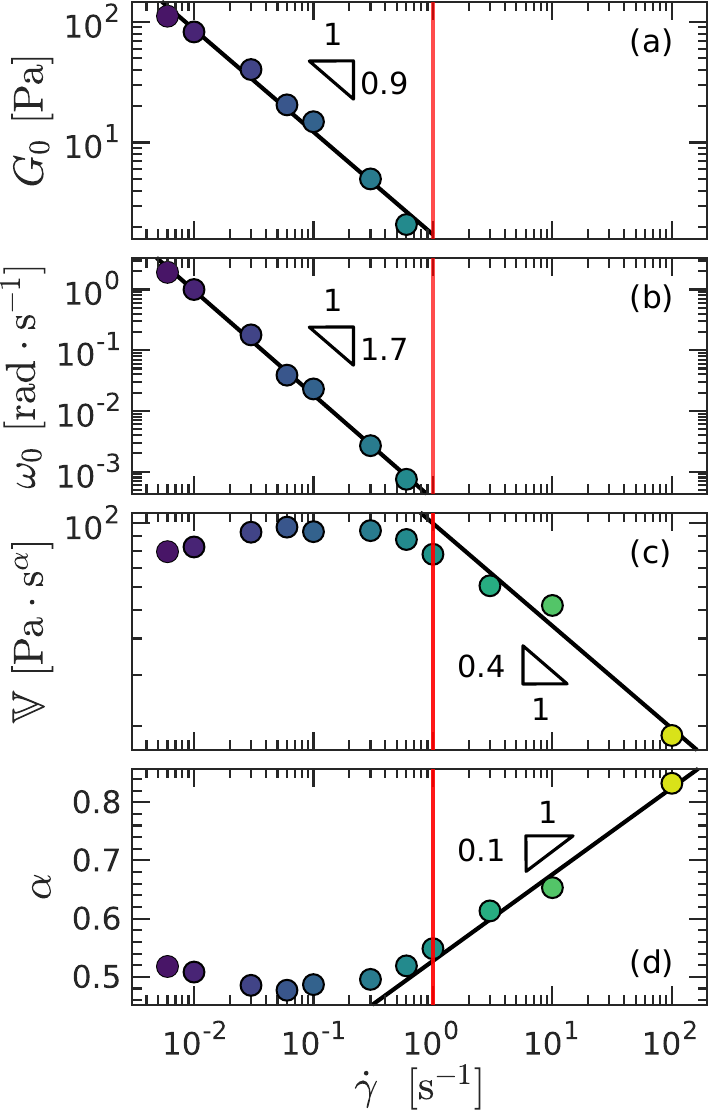}
    \caption{Shear-rate dependence of the Fractional Kelvin-Voigt (FKV) model parameters for an insulating CB-CMC hydrogel ($c_{\rm CMC} = 2\%$ and $x_{\rm CB} = 10\%$). Evolution of (a) the characteristic modulus $G_0$, (b) the characteristic frequency $\omega_0$, (c) the spring-pot quasi-property $\mathbb{V}$, and (d) the fractional exponent $\alpha$ as a function of the applied shear rate $\gpt$. Black lines show the best power-law fits of the data either below or above $\gpt^\star$, with the corresponding exponents indicated in each panel. The vertical red line marks the critical shear rate $\gpt^\star = 1~\rm s^{-1}$.}
    \label{fig:OSP_KVF}
\end{figure}

\begin{figure*}[!t]
    \centering
    \includegraphics[width=0.95\linewidth]{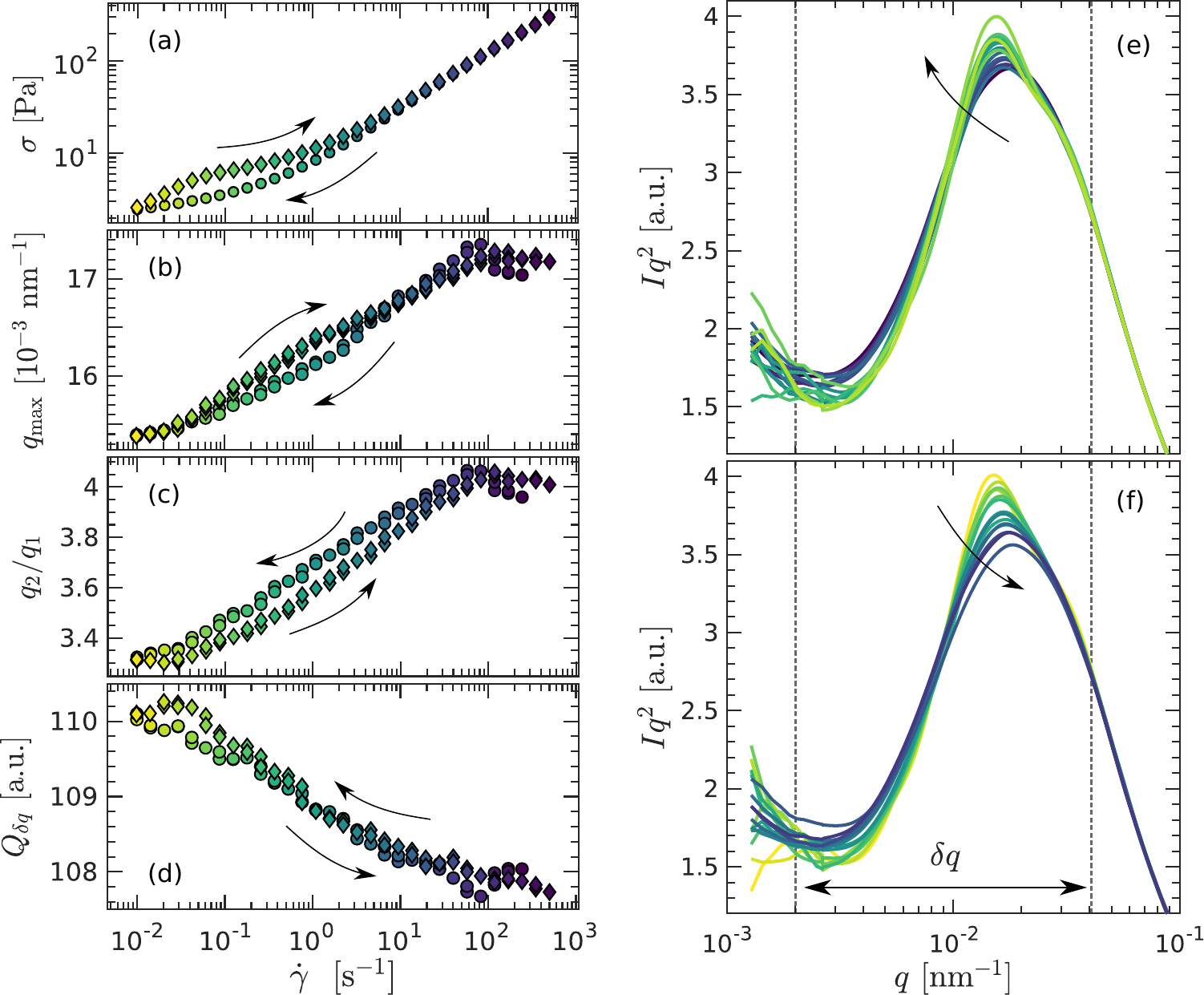}
    \caption{Combined rheology and USAXS measurements during a decreasing and increasing shear rate cycle for an insulating CB-CMC hydrogel ($c_{\rm CMC} = 2\%$ and $x_{\rm CB} = 8\%$). The shear rate is varied stepwise (10~s per step, seven points per decade) from $\gpt=500$ to 0.01~s$^{-1}$ (ramp down, circles) and subsequently from $\gpt= 0.01 $ to 500~s$^{-1}$ (ramp up, diamonds), as indicated by the black arrows. (a) Shear stress $\sigma$. (b) Position $q_{\rm max}$ of the maximum in the Kratky representation $Iq^2(q)$ of the scattering intensity $I$. (c) Peak width quantified by the ratio $q_2 / q_1$, where $q_1$ and $q_2$ denote the positions of the inflection points on either side of $q_{\rm max}$. (d) Integrated area $Q_{\delta q}$ under the Kratky curve between the dashed lines shown in (e,f). (e,f) Kratky representations $Iq^2(q)$ recorded during the decreasing (e) and increasing (f) shear rate sweeps. In (e,f), lighter colors correspond to larger shear rates.}
    \label{fig:saxs_flow_curve_polym}
\end{figure*}

Time-shear-rate superposition has previously been reported for CB and fumed silica dispersions in organic solvents.\cite{Sung:2018} In those systems, only a horizontal frequency shift was required, with the shift factor directly linked to the viscosity, indicating that viscosity controls the dominant relaxation timescale. In contrast, the present superposition involves both horizontal and vertical shifts. Importantly, these two shifts do not reflect independent mechanisms but stem from the single evolution of the elastic scale $G_0$ within the FKV model (Fig.~\ref{fig:OSP_KVF}). For $\gpt < \gpt^\star$, the spring-pot parameters $\mathbb{V}$ and $\alpha$ remain constant, so that the characteristic frequency $\omega_0 = (G_0 / \mathbb{V})^{1/\alpha}$ depends only on $G_0$, and with $\alpha \simeq 0.5$, one directly recovers the scaling $G_0 \sim \omega_0 ^{0.5}$. 
We note that the low shear rate fractional exponent, $\alpha =0.50\pm 0.02$ for the composition of Fig.~\ref{fig:OSP_master_curve} and \ref{fig:OSP_KVF}, is somewhat lower than the value $\alpha=0.62$ found during recovery at rest (see Sec.~\ref{sec:TRMS}) and than the value $\alpha = 0.66$ reported under concentration variations at rest.\cite{Legrand:2023} For the second insulating composition examined [see Fig.~\ref{fig:OSP_KVF2} in Appendix~\ref{appendix:FKVparameters}] the low-shear-rate exponent is instead $\alpha =0.70 \pm 0.02$ in closer agreement with the rest value. This suggests that $\alpha (\gpt \rightarrow0)$ does not extrapolate uniformly onto the quiescent value across compositions, at least within our present resolution, and that a more systematic determination of $\alpha$ and its uncertainty at low shear rates would be needed to establish whether the two limits are quantitatively compatible.

The rescaling observed here, which involves both horizontal and vertical shifts, therefore, has a fundamentally different physical origin from that previously reported and instead reflects a progressive weakening of the physically cross-linked polymer network under shear. 
A plausible microscopic interpretation is that, under continuous shear, CB clusters become less efficient crosslinkers: they dynamically attach and detach from polymer chains, reducing the number of elastically active connections. This effect is enhanced as $\gpt$ increases, leading to the observed decrease of $G_0$. 

The same phenomenology is observed for another insulating composition at $c_{\rm CMC} = 2\%$ and $x_{\rm CB} =8\%$ (see Fig.~\ref{fig:OSP_KVF2} in Appendix~\ref{appendix:FKVparameters}). Below $\gpt^\star$, the FKV parameters exhibit similar shear rate dependence as for the composition discussed in Fig.~\ref{fig:OSP_KVF}. Above $\gpt^\star$, the elastic contribution vanishes, the spring-pot quasi-property $\mathbb{V}$ decreases, and the exponent $\alpha$ increases with $\gpt$, signaling the progressive fluidization of the sample. For $ \gpt \gg \gpt^\star$, the exponent $\alpha\rightarrow 1$, i.e., the spring-pot turns into a regular dashpot, consistent with progressive suppression of the elastic contribution associated with CB-CMC interactions.
The primary difference between the two formulations lies in the value of the critical shear rate $\gpt^\star$, which is significantly lower for the sample with the lowest CB content ($\gpt^\star \simeq 0.3~\rm s^{-1}$ vs.~$1~\rm s^{-1}$). The physical meaning of $\gpt^\star$ remains open: it may reflect either a typical timescale associated with CB-CMC adhesion or its typical energy scale that is overcome by the viscous forces induced by the continuous shear. 
Interestingly, $\gpt^\star$ falls within the range of shear rates over which the rheo-USAXS measurements of Sec.~\ref{sec:rheoTRUSAXS} reveal a shear-dependent reorganization of the spatial distribution of CB clusters within the CMC network (Fig.~\ref{fig:saxs_flow_curve_polym}), raising the possibility that the loss of elasticity in orthogonal superposition and the structural rearrangement seen by USAXS share a common origin. A quantitative way to compare this threshold across compositions would be through a Mason number $\rm Mn=6\pi \eta a^2 \gpt/F_{\rm adh}$, comparing the hydrodynamic drag force on a CB cluster of characteristic size $a$ to the adhesive force $F_{\rm adh}$, binding CB clusters to the CMC network, with $\gpt^\star$ corresponding to $\rm Mn \simeq 1$,\cite{Varga:2018,Varga:2019} although testing this would require additional challenging measurements of $F_{\rm adh}$.

\subsection{Insulating CB-CMC hydrogels: rheo-TRUSAXS}
\label{sec:rheoTRUSAXS}

To complement the orthogonal superposition measurements along the flow curve, we have performed rheo-TRUSAXS experiments during a shear-rate cycle consisting of a decreasing ramp followed by an increasing ramp. 
The insulating CB-CMC hydrogel investigated ($c_{\rm CMC} = 2\%$ and $x_{\rm CB} =8\%$) exhibits the so-called rheological hysteresis\cite{Divoux:2013} [Fig.~\ref{fig:saxs_flow_curve_polym}(a)]: for $\gpt \lesssim 1~\rm s^{-1}$, the stress on the increasing ramp lies above that on the decreasing ramp. 
The simultaneously recorded USAXS data are reported as Kratky plots, $I(q).q^2$ versus $q$, 
for the decreasing and increasing ramps 
in Figs.~\ref{fig:saxs_flow_curve_polym}(e) and \ref{fig:saxs_flow_curve_polym}(f), respectively. 
As previously reported at rest and under large-amplitude oscillatory shear,\cite{Legrand:2026}
the scattering intensity shows a non-monotonic profile, 
with a local maximum at $q=q_{\rm max}$, which combines 
the form factor of individual CB particles at large $q$ with the structure factor associated with the spacing between CB clusters at low $q$.

At large scattering vectors, $q \gtrsim 2.10^{-2}~\rm nm^{-1}$, $Iq^2$ is essentially insensitive to the applied shear rate, consistent with Porod scattering from the surface of the CB particles.\cite{Legrand:2026} In contrast, some shear rate dependence is observed at lower $q$, i.e., it reflects changes in the CB clusters rather than in the particles themselves. Upon decreasing $\gpt$, $q_{\rm max}$ shifts continuously from $q_{\rm max}\simeq 1.7\cdot 10^{-2}~\rm nm^{-1}$ to $1.5\cdot 10^{-2}~\rm nm^{-1}$ [Fig.~\ref{fig:saxs_flow_curve_polym}(b)], while the Kratky peak simultaneously narrows -- quantified by $q_2/q_1$ reported in Fig.~\ref{fig:saxs_flow_curve_polym}(c) -- and grows as shown in Fig.~\ref{fig:saxs_flow_curve_polym}(e). Together, these trends indicate that CB clusters become larger, more widely spaced, and more homogeneous in size as the shear rate is reduced, consistent with their progressive incorporation as physical crosslinks within the CMC network.
Notably, both $q_{\rm max}$ and $q_2/q_1$ display some hysteresis between $0.1~\rm s^{-1}$ and $10~\rm s^{-1}$, over a larger range than the rheological hysteresis observed in Fig.~\ref{fig:saxs_flow_curve_polym}(a).

Following the analysis introduced in our companion study,\cite{Legrand:2026} we also compute the area $Q_{\delta q}$ enclosed by the Kratky plot between $2.10^{-3}$ and $4.10^{-2}~\rm nm^{-1}$ [the window 
$\delta q$ indicated by the vertical dashed lines in Fig.~\ref{fig:saxs_flow_curve_polym}(e) and (f)]. Because scattering at $q \gtrsim 2.10^{-2}~\rm nm^{-1}$ is dominated by Porod scattering from the CB particle surface and is essentially shear-rate independent, any change in $Q_{\delta q}$ necessarily originates from the lower-$q$ end of this window, i.e., from the structure-factor contribution associated with the spacing between CB clusters -- note that if $\delta q$ spanned the entire accessible range, $Q_{\delta q}$ would be constant by construction. This quantity therefore offers a compact way of tracking changes in the low-$q$ part of the spectrum that are not directly captured by $q_{\rm max}$ alone. Here, this area increases monotonically upon decreasing shear rate, indicating a growing contribution of CB clusters to mesoscale structural correlations. In contrast to the more pronounced hysteresis observed for $q_{\rm max}$ and the peak width $q_2/q_1$, the hysteresis of $Q_{\delta q}$ remains weak and is restricted to low shear rates $\gpt \lesssim 0.1~\rm s^{-1}$, where the hysteresis observed in the $\sigma(\gpt)$ response is the largest.  

This indicates that the rheological hysteresis primarily originates from a shear-history-dependent reorganization of the CB clusters --their mean spacing and spatial heterogeneity-- rather than from a substantial transfer of CB particles into or out of correlated structures: the area $Q_{\delta q}$ varies only weakly (from about 108 to 110~a.u.) compared with the pronounced changes in $q_{\rm max}$ and $q_2/q_1$, consistent with a redistribution of existing clusters rather than their large-scale formation or breakage. 
The weak hysteresis of $Q_{\delta q}$ nonetheless suggests that a small fraction of CB clusters is reversibly recruited or released as crosslinks within the CMC network, consistent with the physical and dynamic nature of CB-polymer interactions.

By contrast, none of the USAXS metrics ($q_{\rm max}$, $q_2/q_1$, $Q_{\delta q}$) show significant variations for a conductive CB-CMC hydrogel subjected to the same shear-rate cycle (Fig.~\ref{fig:saxs_flow_curve_coll}, Appendix~\ref{appendix:C}). This indicates that shear-induced restructuring of the percolated CB network --likely rupture into large-scale aggregates or fractures-- occurs at length scales beyond the USAXS window probed here.

\section{Discussion and Conclusion}
 \label{sec:discussion}

Let us now summarize the main results of this work. We have shown that the rheological response of CB-CMC hydrogels depends on their underlying microstructure, which is controlled by the CMC-to-CB ratio. Conductive and insulating hydrogels display markedly different aging dynamics despite sharing the same constituents. Conductive gels recover through a logarithmic increase of elasticity characteristic of percolated colloidal networks, whereas insulating gels exhibit the coexistence of logarithmic and power-law recovery associated with distinct relaxation mechanisms. These different aging behaviors are accompanied by distinct electrical signatures, emphasizing the intimate connection between mechanical and electrical relaxation in these hybrid materials. Building on these observations, we demonstrate that the evolution of the viscoelastic spectrum of insulating gels obeys generalized superposition principles both during aging and under steady shear. 

A first key result of this work is that logarithmic aging is observed in both conductive and insulating CB-CMC hydrogels despite their fundamentally different microstructures. In conductive hydrogels, the recovery rate depends only on the carbon black content and is essentially insensitive to the amount of polymer, indicating that aging is governed by the progressive evolution of the percolated CB network. In contrast, rheo-USAXS measurements show that the logarithmic recovery of insulating hydrogels occurs while the average CB structure remains essentially unchanged. The increase in elasticity during this stage must therefore originate from local relaxation processes within the polymer network or at the polymer-particle interfaces rather than from large-scale rearrangements of the CB crosslinkers. 
This interpretation is supported by our electrical impedance signatures: as the elastic modulus of the insulating gel increases through the power-law contribution to aging, the increasing viscosity should lower the ionic diffusion coefficient and hence the DC-like conductivity, consistent with the modest decrease of $\epsilon''$ at low frequency.
This comparison between conductive and insulating CB-CMC hydrogels suggests that logarithmic aging should not be viewed as the signature of a specific microscopic mechanism but rather as the generic rheological manifestation of progressively slower relaxation toward deeper metastable states, irrespective of the structural degrees of freedom involved.\cite{Joshi:2015,Boettcher:2018} 
Only insulating hydrogels exhibit a second, power-law stiffening regime, which coincides with a measurable shift of the USAXS peak toward larger scattering vectors. This observation suggests that, once local polymer relaxation has largely taken place, the recovery proceeds through a slower densification of the CB-mediated crosslinks, leading to additional reinforcement of the polymer network.

A second important outcome of this work concerns the comparison between the different superposition principles observed in insulating CB-CMC hydrogels. During recovery after flow cessation, the viscoelastic spectra obey a time-aging superposition that closely resembles the time-concentration superposition previously reported for the same material.\cite{Legrand:2023} In both cases, the spectra are well described by a Fractional Kelvin-Voigt model with a similar fractional exponent ($\alpha\simeq 0.62$ in the present work versus $\alpha\simeq 2/3$ under concentration variations), and the characteristic modulus and frequency remain proportional ($G_0\propto\omega_0$). This close agreement indicates that both aging at rest and increasing CB content at a fixed point in time lead to the progressive buildup of crosslinks between CB and CMC, effectively controlling the same underlying network connectivity. In this sense, the waiting time can be mapped onto an increasing effective CB concentration, revealing a generalized time-aging-concentration superposition principle for insulating CB-CMC hydrogels, while preserving the overall spectrum of relaxation modes.

By contrast, the viscoelastic spectra measured under steady shear by orthogonal superposition obey a different scaling. Although they also collapse onto a single master curve and remain well described by the same Fractional Kelvin-Voigt framework, the fractional exponent decreases to $\alpha\simeq0.5$, while the shift factors satisfy $G_0\propto\omega_0^{0.5}$. Remarkably, the equality between the fractional exponent and the scaling exponent relating $G_0$ and $\omega_0$ is precisely the behavior predicted for near-critical gels,\cite{Martin:1988,Martin:1989,Costanzo:2020} where both quantities follow critical scaling with the distance to the gel point owing to the self-similar structure of the stress-bearing network. This correspondence suggests that moderate steady shear continuously reduces the connectivity of transient CB junctions --likely governed by the dynamic adsorption and desorption kinetics of the stabilizing CMC chains under flow-- in a manner analogous to moving away from the gel state toward a shear-induced critical regime. Thus, while quiescent aging drives the system deeper into the gel state like an effective concentration jump, steady shear acts as an opposing lever that tunes the network back toward its critical threshold.
Above a critical shear rate $\dot \gamma^*$, the rate of structural disruption exceeds the rate of bond reformation, leading to a breakdown of the gel network and a transition to a flowing, non-gelled state, thereby clarifying the physical origin of this critical shear threshold.

To conclude, this work illustrates how combining rheology, electrical impedance spectroscopy, and time-resolved X-ray scattering provides complementary access to the evolution of soft colloidal networks over multiple length scales. Extending this multimodal approach to other transient polymer-particle networks should help establish more general connections between microscopic restructuring, aging dynamics, and nonlinear rheology.

\begin{acknowledgments}
The authors thank W.~Ch\`evremont and T.~Narayanan for helpful discussions and technical support during the Rheo-USAXS experiments, and R.~Cardinaels, A.~Colin, P.~Coussot, A.-C. Genix, J.J.~Richards, and J.~Vermant for fruitful discussions.
The ESRF is acknowledged for providing
synchrotron beamtime on ID02 (proposal SC-5515). 
This work was supported by the LABEX iMUST of the University of Lyon (ANR-10-LABX-0064), created within the ``Plan France 2030" set up by the French government and managed by the French National Research Agency (ANR).
\end{acknowledgments}


\appendix

\begin{figure}[t]
    \centering
    \includegraphics[width=0.95\columnwidth]{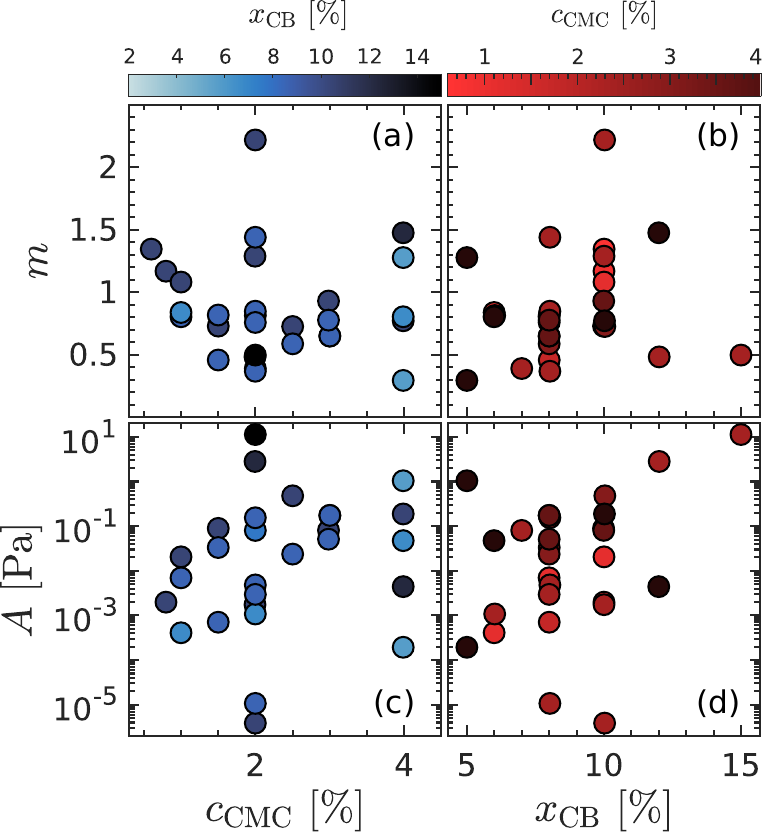}
    \caption{Parameters of the power law component in the aging dynamics from Eq.~\eqref{eq:aging} in the insulating regime. Exponent $m$ as a function of $c_{\rm CMC}$ (a) and $x_{\rm CB}$ (b). Prefactor $A$ as a function of $c_{\rm CMC}$ (c) and $x_{\rm CB}$ (d). When the x-axis is $c_{\rm CMC}$, the color code for $x_{\rm CB}$ according to the colorbar given in the left column (lighter color corresponds to lower CB content). When the x-axis is $x_{\rm CB}$, the color code is for $c_{\rm CMC}$ according to the colorbar given in the right column (lighter color corresponds to lower CMC concentration). } \label{Fig:paramters_TS_powerlaw}
\end{figure}

\begin{figure}[b]
    \centering
    \includegraphics[width=0.95\linewidth]{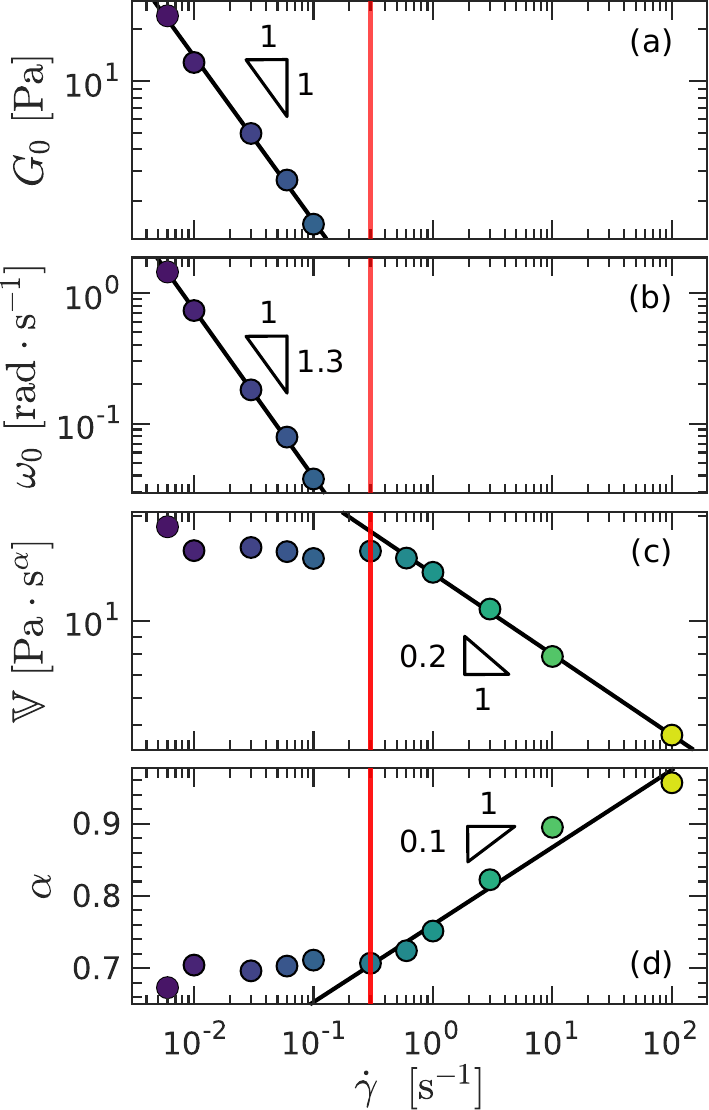}
    \caption{Shear-rate dependence of the Fractional Kelvin-Voigt (FKV) model parameters for an insulating CB-CMC hydrogel ($c_{\rm CMC} = 2\%$ and $x_{\rm CB} = 8\%$). Evolution of (a) the characteristic modulus $G_0$, (b) the characteristic frequency $\omega_0$, (c) the spring-pot quasi-property $\mathbb{V}$, and (d) the fractional exponent $\alpha$ as a function of the applied shear rate $\gpt$. Black lines show the best power-law fits of the data either below or above $\gpt^\star$, with the corresponding exponents indicated in each panel. The vertical red line marks the critical shear rate $\gpt^\star = 0.3~\rm s^{-1}$.}
    \label{fig:OSP_KVF2}
\end{figure}

\begin{figure*}[!ht]
    \centering
    \includegraphics[width=0.95\linewidth]{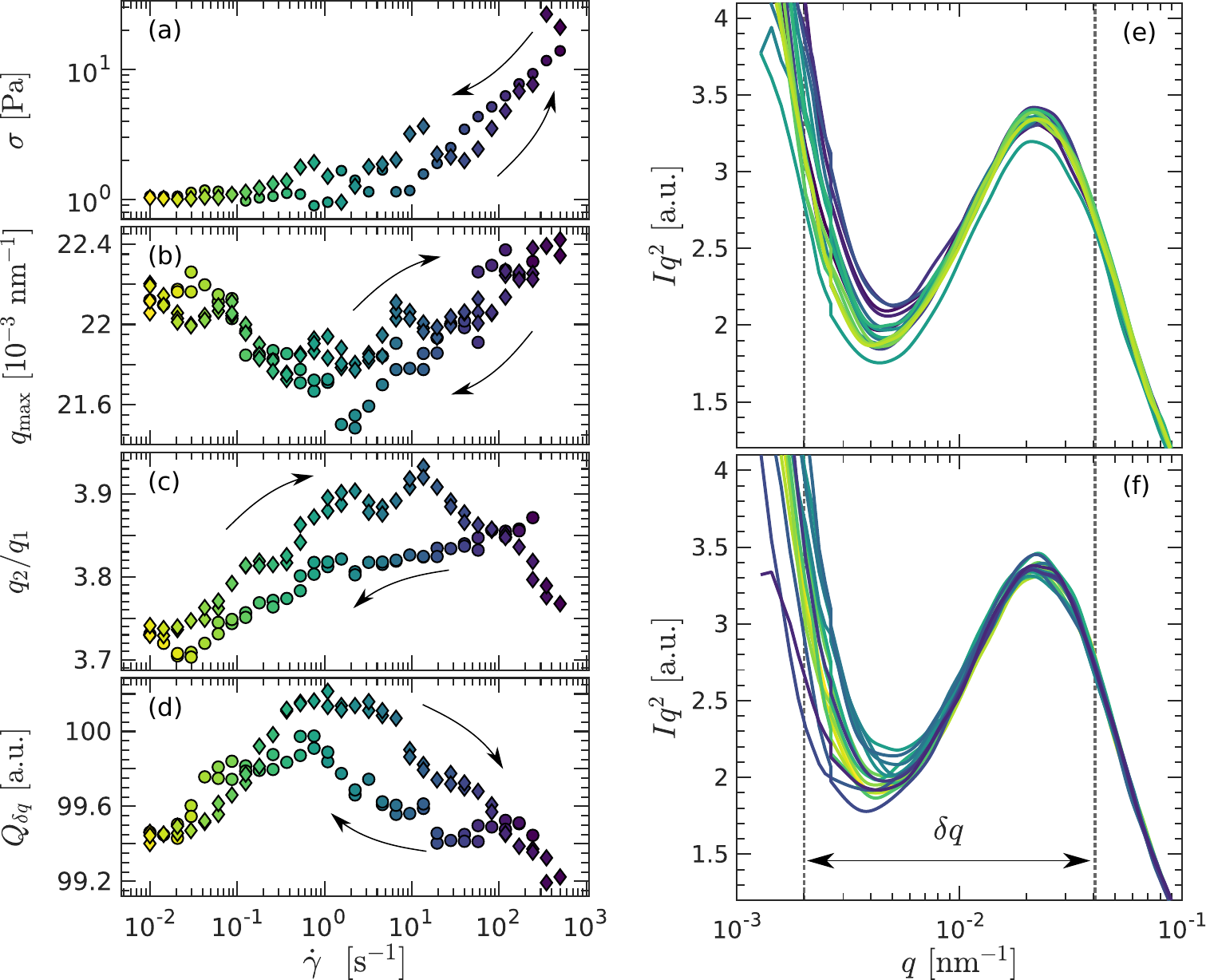}
    \caption{Combined rheology and USAXS measurements during a decreasing and increasing shear rate cycle for a conductive CB-CMC hydrogel ($c_{\rm CMC} = 0.01\%$ and $x_{\rm CB} = 6\%$). The shear rate is varied stepwise (10~s per step, seven points per decade) from $\gpt=500$ to 0.01~s$^{-1}$ (ramp down, circles) and subsequently from $\gpt= 0.01 $ to 500~s$^{-1}$ (ramp up, diamonds), as indicated by the black arrows. (a) Shear stress $\sigma$. (b) Position $q_{\rm max}$ of the maximum in the Kratky representation $Iq^2(q)$ of the scattering intensity $I$. (c) Peak width quantified by the ratio $q_2 / q_1$, where $q_1$ and $q_2$ denote the positions of the inflection points on either side of $q_{\rm max}$. (d) Integrated area under the Kratky curve between the dashed lines shown in (e,f). (e,f) Kratky representations $Iq^2(q)$ recorded during the decreasing (e) and increasing (f) shear rate sweeps. In (e,f), lighter colors correspond to larger shear rates.}
    \label{fig:saxs_flow_curve_coll}
\end{figure*}

\section{Fit parameters of physical aging in insulating CB-CMC hydrogels}
\label{appendix:Fitaging}

In section~\ref{sec:aginginsulating}, we have shown that the recovery of insulating CB-CMC gels is well described by the sum of a logarithmic and a power-law contribution [see Eq.~\eqref{eq:aging}]. Figure~\ref{Fig:paramters_TS_powerlaw} displays the dependence of the fit parameters $A$ and $m$, which characterize the amplitude and exponent of the power-law contribution, respectively, with respect to the CB and CMC content. Neither the amplitude $A$ nor the exponent $m$ exhibit a systematic dependence on the CB or the CMC content.

\section{Impact of shear on the FKV model parameters of an insulating CB-CMC hydrogel at $c_{\rm CMC}=2\%$ \ $x_{\rm CB}=8\%$}
\label{appendix:FKVparameters}

In Figure~\ref{fig:OSP_KVF2}, we report the impact of shear on the FKV model parameters of an insulating CB-CMC hydrogel at $c_{\rm CMC}=2\%$ and $x_{\rm CB}=8\%$, to be compared with similar data reported in Fig.~\ref{fig:OSP_KVF} in the main text for an insulating hydrogel with a larger CB content ($c_{\rm CMC}=2\%$ \& $x_{\rm CB}=10\%$). We observe similar trends, with one striking difference: the critical shear rate $\dot \gamma^*$ beyond which the elastic properties vanish is smaller for $x_{\rm CB}=8\%$, which suggests that $\gamma^*$ increases with increasing CB content.

\section{Rheo-TRUSAXS on a conductive CB-CMC hydrogel}
\label{appendix:C}
Figure~\ref{fig:saxs_flow_curve_coll} reports the result of rheo-TRUSAXS measurements performed during decreasing and increasing shear rate sweeps for a conductive CB-CMC hydrogel. The figure follows the same layout as Fig.~\ref{fig:saxs_flow_curve_polym} in the main text. In contrast to the insulating hydrogels, the scattering intensity remains essentially unchanged throughout the shear rate cycle. In particular, no significant changes are observed in either
the position of the maximum of the Kratky representation, $Iq^2$, or its width. The same behavior is observed during both the decreasing and increasing shear rate sweeps, suggesting that the shear-induced microstructural changes responsible for the rheological response occur at length scales beyond the USAXS window, i.e., larger than a few micrometers.


\clearpage


%

\end{document}